\documentclass[preprint,3p,12pt]{elsarticle}
\usepackage{mathrsfs}
\usepackage{amsmath}
\usepackage{stmaryrd}
\usepackage{bbding}
\usepackage{dcolumn}
\usepackage{graphicx}
\usepackage{amsfonts}
\usepackage{amssymb}
\usepackage{psfrag}
\usepackage{wrapfig}
\usepackage{subfigure}
\usepackage{makeidx}
\usepackage{bm}
\usepackage{epsf}
\usepackage{epsfig}
\usepackage{setspace}
\usepackage{graphicx}
\usepackage{epstopdf}
\usepackage{psfrag}
\usepackage{subfigure}
\usepackage{booktabs}
\usepackage{color}
\usepackage{comment}

\begin{document}
	
\title{Unified gas-kinetic wave-particle method for gas-particle two phase flow from dilute to dense solid-particle limit}
	
\author[HKUST1]{Xiaojian Yang}
\ead{xyangbm@connect.ust.hk}
	
\author[HKUST1]{Wei Shyy}
\ead{weishyy@ust.hk}
	
\author[HKUST1,HKUST2,HKUST3]{Kun Xu\corref{cor1}}
\ead{makxu@ust.hk}	
	
\address[HKUST1]{Department of Mechanical and Aerospace Engineering, Hong Kong University of Science and Technology, Clear Water Bay, Kowloon, Hong Kong, China}
\address[HKUST2]{Department of Mathematics, Hong Kong University of Science and Technology, Clear Water Bay, Kowloon, Hong Kong, China}
\address[HKUST3]{Shenzhen Research Institute, Hong Kong University of Science and Technology, Shenzhen, China}
\cortext[cor1]{Corresponding author}

\begin{abstract}
In this paper, a unified framework for particulate two-phase flow will be presented with a wide range of solid-particle concentration from dilute to dense limit.
The two phase flow is simulated by two coupled flow solvers, i.e.,  the gas-kinetic scheme (GKS) for the gas phase and unified gas-kinetic wave-particle method (UGKWP) for the solid-particle phase. The GKS is a second-order Navier-Stokes flow solver for the continuum gas flow.
The UGKWP is a multiscale method for all flow regimes. The wave and particle decomposition in UGKWP depends on the cell's Knudsen number (Kn).
At a small Kn number, the high concentrated solid particle phase will be modeled by the Eulerian hydrodynamic wave due to the intensive particle-particle collisions, same as the fluid model. At a large Kn number, the dilute solid particle will be sampled and followed by the Lagrangian particle formulation to capture the non-equilibrium transport.
In the transition regime, the distribution and evolution of particle and wave in UGKWP are controlled by the local Kn number with a smooth transition between the above limits. The distribution of solid particles in UGKWP is composed of analytical function and discrete particle, where both condensed and dilute phases can be automatically captured in the most efficient way.
In the current scheme, the two phase model improves the previous one in all following aspects: drag force model for different solid particle concentrations; the frictional pressure in inter-particle contacts at high solid-particle concentration; a flux limiting model to avoid solid particles' over-packing; additional non-conservative nozzle and work terms in the governing equation for the gas phase to reflect the local variation of solid volume fraction.
Besides, the inter-particle collisions have been refined numerically for the dense particle phase flow through
 the discretization of the collision term and numerical flux function.
 The improved method has been applied to gas-particle system with a wide range of solid-particle concentrations.
The numerical scheme is tested in a series of typical gas-particle two-phase problems, including the interaction of shock wave with solid particle layer, horizontal pneumatic conveying, bubble formation and particle cluster phenomena in the fluidized bed. The results validate the accuracy and reliability of the proposed method for gas-particle flow.
\end{abstract}


\maketitle

\section{Introduction}
Gas-particle two-phase flow is very common in natural phenomena, e.g., sand storms, volcano eruption, and many engineering industries, e.g., petroleum industry, chemical industry, energy industry. Numerical simulation is a powerful tool to study the gas-particle two-phase flow, and many numerical methods have been developed to accurately and efficiently capture the complex physics of gas-particle flow \cite{Gasparticle-review-multiscale-tsuji2007multi, Gasparticle-review-van2008numerical, Gasparticle-review-zhongwenqi-zhong2016cfd, Gasparticle-review-Ge2017discrete, Gasparticle-review-WangJunwu2020continuum}.

Generally two approaches, Eulerian-Eulerian (EE) approach and Eulerian-Lagrangian (EL) approach, are widely employed, and the difference of this classification is based on the treatment of particle phase.
In the EE approach, the particle phase is assumed as a continuum media, and hydrodynamic equations are employed for the evolution of particle flow \cite{Gasparticle-Abgrall-saurel1999multiphase, Gasparticle-TFM-compressible-houim2016multiphase}. EE approach is also called two fluid model (TFM). One representative EE approach is kinetic theory-based granular flow (KTGF), which is based on the similarity in the modeling of solid particle and the molecule in gas \cite{Gasparticle-KTGF-ding1990bubbling, Gasparticle-book-luhuilin2021computational}.
In EL approach, all individual solid particles or particle parcels composed of a group of solid particles with the same properties, will be tracked according to Newton's law of motion in the simulation \cite{ Gasparticle-DEM-review-guo-curtis-2015discrete}. Some typical EL approaches are discrete element method (DEM) \cite{Gasparticle-DEM-cundall1979discrete, Gasparticle-DEM-tsuji1993discrete, Gasparticle-DEM-review-guo-curtis-2015discrete}, coarse-grained particle method (CGPM) \cite{Gasparticle-coarse-grained-DEM-sakai2009large, Gasparticle-coarse-grained-EMMS-DPM-gewei-lu2016computer}, multiphase particle-in-cell (MP-PIC) \cite{Gasparticle-PIC-andrews1996multiphase, Gasparticle-PIC-rourke-2009model, Gasparticle-PIC-rourke-2012inclusion}, etc.
In terms of the consideration of flow physics, the choice of EE or EL depends on the local Knudsen ($Kn$) number of particle flow.
Similar to gas, the $Kn$ number of disperse phase can be defined as the ratio of mean free path (MFP) of solid particles over characteristic length scale \cite{Gasparticle-momentmethod-Fox2013computational}. When $Kn$ number is very small with sufficient inter-particle collisions,
the solid particle phase can be assumed as a continuum medium, and the EE approach can be appropriately used for the gas-particle system.
On the contrary, when $Kn$ number is large, individual particle transport becomes important and the solid phase stays in a non-equilibrium state. So, the EL approach is a preferred choice. The disadvantage of EL approach is the high computational cost due to the particle trajectory tracking for all individual particles or parcels, especially in the dense solid-particle flow\cite{Gasparticle-review-van2008numerical}.
Theoretically, EL approach can be used when $Kn$ number is small as long as the computation cost is affordable.
For the EE approach, it will difficult to give an accurate prediction when $Kn$ number of particle phase is large,
because EE approach cannot capture non-equilibrium physics of solid particles, such as particle trajectory crossing (PTC) phenomenon \cite{Gasparticle-review-balachandar2010turbulent, Gasparticle-momentmethod-Fox2013computational}.
Based on the features of EE and EL approach, many studies focus on the hybrid method, coupling Eulerian and Lagrangian approach together for solid particle phase, to maintain both the accuracy and computation efficiency \cite{Gasparticle-hybrid-EL-pialat2007hybrid, Gasparticle-hybrid-dynamic-multiscale-method-chen-wangjunwu-2017dynamic, Gasparticle-hybrid-KTGF-DEM-zhang-luhuilin-2019modified, Gasparticle-hybrid-EL-panchal2021hybrid}. In the hybrid method, it is a challenge to define an accurate and reliable criterion for the smooth transition between the Eulerian and Lagrangian approaches for disperse phase.
In addition, some other methods are proposed and used for the gas-particle flow, such as direct numerical simulation (DNS) \cite{Gasparticle-DNS-luokun-li2016direct, Gasparticle-DNS-gewei-liu2017meso}, unified gas kinetic scheme (UGKS) \cite{UGKS-xu2010unified, UGKS-gas-particle-liu2019unified}, unified gas kinetic particle method (UGKP) \cite{KP-gasparticle-wangzhao-wang2020unified}, discrete unified gas kinetic scheme (DUGKS) \cite{Gasparticle-DUGKS-immersed-boundary-guozhaoli-tao2018combined}, method of moment (MOM) \cite{Gasparticle-MOM-Fox-desjardins2008quadrature, Gasparticle-momentmethod-Fox2013computational}, direct simulation Monte Carlo (DSMC) \cite{Gasparticle-DSMC-bird1976molecular}, material point method (MPM) \cite{Gasparticle-MPM-baumgarten2019general}, smooth particle hydrodynamics (SPH) \cite{Gasparticle-SPH-GeWei-deng2013two}, hybrid coarse-grain DEM and resolved DEM \cite{Gasparticle-hybrid-DEM-coarse-grained-DEM-queteschiner2018coupling}, hybrid finite-volume-particle method \cite{Gasparticle-dusty-hybrid-finite-volume-particle-chertock2017hybrid}, etc.

In recent years, unified gas-kinetic scheme (UGKS) has been developed for rarefied and continuum flow simulation \cite{UGKS-xu2010unified, UGKS-book-xu2014direct}. Based on the direct modeling on the cell's Knudsen number, i.e., $Kn_c = \tau/\Delta t$ with particle collision time $\tau$ over numerical time step $\Delta t$, UGKS recovers multiscale transport in flow regimes through a smooth connecting between $e^{-1/Kn_c}$ weighted equilibrium flow evolution and the rest  $(1-e^{-1/Kn_c})$ particle free transport, and the NS solution is automatically obtained at small $Kn_c$. After the success of the UGKS for the gas flow, the method has been further extended to other multiscale transports,
such as radiative heat transfer, neutron transport, plasma, particulate flow, etc \cite{UGKS-radiative-sun2015asymptotic, UGKS-neutron-tan2020,UGKS-plasma-liu2017unified, UGKS-gas-particle-liu2019unified}. A particle-based UGKS, which is named unified gas-kinetic particle (UGKP) method, was developed subsequently using stochastic particles to follow the evolution of gas distribution function \cite{WP-first-liu2020unified, WP-second-zhu-unstructured-mesh-zhu2019unified}.
In UGKP, the sampled particles can be divided into two categories: collisionless (free transport) particle and collisional particle within each time step. The collisional particles will be eliminated in the evolution and get re-sampled from the equilibrium state at the beginning of the next time step. As a result, only the collisionless particles are fully tracked in the whole time step in UGKP.
Furthermore, it is realized that a proportion $e^{-1/Kn_c}$ of re-sampled particles from the equilibrium state at the beginning of next time step in UGKP will get collision and be eliminated again within the next time step. Actually, the contribution from these re-sampled collisional particles to flux function in the finite volume UGKP can be evaluated analytically. As a result, the collisional particles don't need to be re-sampled at all, and can be followed analytically through a wave representation in the upgraded unified gas-kinetic wave-particle (UGKWP) method \cite{WP-first-liu2020unified, WP-second-zhu-unstructured-mesh-zhu2019unified, WP-3D-chen2020three, WP-sample-xu2021modeling}. In UGKWP, wave and particle are coupled together in the evolution, and only free transport particles are basically tracked to capture the non-equilibrium flow physics. Therefore, UGKWP becomes a hydrodynamic flow solver in the continuum flow regime due to the absence of particles and goes to a particle method in the highly rarefied regime. UGKWP can present an optimized approach to capture
multiscale transport efficiently using the combination of wave and particle.
In the continuum flow regime, UGKWP will automatically get back to the gas-kinetic scheme (GKS), which is
a kinetic theory-based Navier-Stokes solver \cite{GKS-2001, GKS-lecture, CompactGKS-ji2020-unstructured, CompactGKS-zhao2019-8th-order, GKS-turbulence-implicitHGKS-Cao2019}. Besides gas flow, UGKWP has also been used in the study of radiative transfer, plasma, and two phase flow
\cite{UGKWP-Li2020,UGKS-Liu2021-AIA,WP-six-gas-particle-yang2021unified}.
The special wave and particle decomposition in UGKWP makes it suitable for the simulation of both dense (wave) and dilute solid-particle (particle) phase easily.

For the particulate two phase flow, the gas phase will be followed by the GKS and solid-particle phase by the UGKWP, and final scheme is called GKS-UGKWP for convenience. For the dilute monodisperse particulate flow, a previous GKS-UGKWP has been developed \cite{WP-six-gas-particle-yang2021unified}. Based on the UGKWP for the solid-particle phase, the sampled particles depends on the local particle's cell's Knudsen number. When $Kn_c$ is extremely small for dense particle distribution, no particle will be sampled in UGKWP and UGKWP reduces to the hydrodynamic flow solver. As a result, the GKS-UGKWP automatically becomes an EE approach. When $Kn$ number is extremely large, only particle evolution in UGKWP will be tracked and the corresponding GKS-UGKWP becomes an EL approach. For the intermediate $Kn_c$ number, both EE and EL formulation will be coupled in each cell according to $Kn_c$ in the evolution of the particulate flow.
In this paper, more realistic model will be implemented in GKS-UGKWP for the two-phase flow simulation.

Based on solid volume fraction $\epsilon_{s}$, the particulate flow is usually divided into dilute flow with $\epsilon_{s} \le \epsilon_{s}^{*}$ and dense flow $\epsilon_{s} > \epsilon_{s}^{*}$, and one of the choices of $\epsilon_{s}^{*}$ is 0.001 \cite{Gasparticle-review-van2006multiscale}. However, the solid volume fraction is not necessarily a reliable indicator showing the importance of particle-particle collision, but the Kundsen number is a suitable indicator \cite{Gasparticle-momentmethod-Fox2013computational}. Generally, the inter-particle collision is (much possibly but not necessarily) more frequent in dense flow than dilute one due to a large number of solid particles. Therefore the particle-particle collision usually plays a significant role in the solid phase evolution of dense phase, and it cannot be neglected in the numerical simulation aiming to accurately recover the real flow physics.
The influence of inter-particle collision is considered and modeled differently in numerical methods. For example, in MP-PIC, an inter-particle stress term models the effect of particle-particle collision, but it can only simulate the particulate flow with solid concentration $\epsilon_{s}<0.05$, which cannot be very high \cite{Gasparticle-PIC-andrews1996multiphase}. With the modification of collision term, the improved MP-PIC can be used for dense particle flow with high concentration \cite{Gasparticle-PIC-rourke-2009model, Gasparticle-PIC-rourke-2012inclusion}.  In DEM, both soft-sphere model and hard-sphere model can be used to calculate the influence of inter-particle collision \cite{Gasparticle-DEM-cundall1979discrete, Gasparticle-DEM-hard-sphere-kuipers-hoomans1996discrete, Gasparticle-review-multiscale-tsuji2007multi}. In UGKWP, the collision effect is explicitly included in the collision term of the kinetic equation for modeling the evolution process from
local non-equilibrium  to equilibrium state \cite{Gasparticle-KTGF-lun1984kinetic, WP-first-liu2020unified}.
For the numerical simulation of dense solid-particle flow, a challenge is the existence of non-conservative ``nozzle term" in momentum equation and correspondingly $pDV$ work term in energy equation for the gas flow, which is similar to $pDV$ term in the quasi-one-dimensional gas nozzle flow equation \cite{Gasparticle-TFM-compressible-houim2016multiphase}. If these terms were not solved correctly, un-physical fluctuations of pressure and flow field would be generated, especially in the flow zone with a steep interface of solid-phase concentration \cite{Gasparticle-Abgrall-saurel1999multiphase, Gasparticle-TFM-compressible-houim2016multiphase}.
When the solid phase approaches to a packing limit, the effect of enduring particle-particle contact and friction, modeled by the solid frictional pressure term, has to be considered \cite{Gasparticle-KTGF-pressure-friction-johnson1987frictional, Gasparticle-pressure-friction-srivastava2003analysis, Gasparticle-pressure-friction-schneiderbauer2012comprehensive}. Also, the introduction of frictional pressure can avoid the solid particles' over-assembling due to the dramatically increased value when the solid volume fraction approaches its maximum limiting value \cite{Gasparticle-KTGF-pressure-friction-johnson1987frictional, Gasparticle-TFM-compressible-houim2016multiphase}.
Particulate flow with high concentration is very common in practical engineering problems, such as fluidized bed, pneumatic conveying, etc \cite{Gasparticle-book-fan1999principles, Gasparticle-book-luhuilin2021computational}. Therefore in this paper, the previously developed GKS-UGKWP for dilute flow is extended to dense gas-particle flow. The GKS-UGKWP is further developed for gas-particle two-phase flow with a wide range of volume fraction from very dilute flow to dense solid-particle phase.

This paper is organized as follows. Section 2 introduces the governing equations for particle phase and the UGKWP method. Section 3 is the governing equations for gas phase and the GKS method. Section 4 introduces the numerical experiments. The last section is the conclusion.

\section{UGKWP for solid-particle phase}
\subsection{Governing equation for particle phase}
The evolution of particle phase is govern by the following kinetic equation,
\begin{gather}\label{particle phase kinetic equ}
\frac{\partial f_{s}}{\partial t}
+ \nabla_x \cdot \left(\textbf{u}f_{s}\right)
+ \nabla_u \cdot \left(\textbf{a}f_{s}\right)
= \frac{g_{s}-f_{s}}{\tau_{s}},
\end{gather}
where $\textbf{u}$ is the particle velocity, $\textbf{a}$ is the particle acceleration caused by the external force, $\nabla_x$ is the divergence operator with respect to space, $\nabla_u$ is the divergence operator with respect to velocity, $\tau_s$ is the relaxation time for the particle phase, $f_{s}$ is the distribution function of particle phase, and $g_{s}$ is the associated equilibrium distribution, which can be written as,
\begin{gather*}
g_{s}=\epsilon_s\rho_s\left(\frac{\lambda_s}{\pi}\right)^{\frac{3}{2}}e^{-\lambda_s \left[(\textbf{u}-\textbf{U}_s)^2\right]},
\end{gather*}
where $\epsilon_s$ is the volume fraction of particle phase, $\rho_s$ is the material density of particle phase, $\lambda_s$ is the value relevant to the granular temperature $T_s$ with $\lambda_s = \frac{m_s}{2k_BT_s}$, $m_s=\rho_s \frac{4}{3}\pi\left(\frac{d_s}{2}\right)^3$ is the mass of one particle, $d_s$ is the diameter of solid particle, and $\textbf{U}_s$ is the macroscopic velocity of particle phase. The sum of kinetic and thermal energy for colliding particle may not be conserved due to the inelastic collision between particles. Therefore the collision term in Eq.\eqref{particle phase kinetic equ} should satisfy the following compatibility condition \cite{UGKS-gas-particle-liu2019unified},
\begin{equation}\label{particle phase compatibility condition}
\frac{1}{\tau_s} \int g_s \bm{\psi} \text{d}\Xi=
\frac{1}{\tau_s} \int f_s \bm{\psi}' \text{d}\Xi,
\end{equation}
where $\bm{\psi}=\left(1,\textbf{u},\displaystyle \frac{1}{2}\textbf{u}^2\right)^T$ and $\bm{\psi}'=\left(1,\textbf{u},\displaystyle \frac{1}{2}\textbf{u}^2+\frac{r^2-1}{2}\left(\textbf{u}-\textbf{U}_s\right)^2\right)^T$. The lost energy due to inelastic collision in 3D can be written as,
\begin{gather*}
Q_{loss} = \frac{\left(1-r^2\right)3p_s}{2},
\end{gather*}
where, $r\in\left[0,1\right]$ is the restitution coefficient, determining the percentage of lost energy in inelastic collision. While $r=1$ means no energy loss (elastic collision), $r=0$ refers to total loss of all internal energy of particle phase $\epsilon_s\rho_se_s =\frac{3}{2}p_s$ with $p_s=\frac{\epsilon_s\rho_s}{2\lambda_s}$.

The particle acceleration $\textbf{a}$ is determined by the external force. In this paper, the drag force $\textbf{D}$, the buoyancy force $\textbf{F}_b$, and gravity $\textbf{G}$ are considered. $\textbf{D}$ and $\textbf{F}_b$ are inter-phase force, standing for the force applied on the solid particles by gas flow. The general form of drag force can be written as,
\begin{gather}\label{drag force model}
\textbf{D} = \frac{m_s}{\tau_{st}}\left(\textbf{U}_g-\textbf{u}\right),
\end{gather}
where $\textbf{U}_g$ is the macroscopic velocity of gas phase, and $\tau_{st}$ is the particle internal response time. Many studies have been conducted on the drag force model to give an accurate prediction for the drag under different solid concentrations. In this paper, the drag force model proposed by Gidaspow is employed to determine $\tau_{st}$ \cite{Gasparticle-book-gidaspow1994multiphase},
\begin{equation}\label{taust equation}
\tau_{st}=
\left\{\begin{aligned}
&\frac{4}{3}\frac{\rho_s d_s}{\rho_g|\textbf{U}_g-\textbf{u}|C_d} \epsilon_{g}^{2.65},  & \epsilon_{g}>0.8, \\
&\frac{1}{150\frac{\epsilon_{s} \mu_g}{\epsilon_{g} \rho_{s} d_s^2} + 1.75\frac{\rho_g |\textbf{U}_g - \textbf{u}|}{\rho_{s} d_s}},  & \epsilon_{g} \le 0.8,
\end{aligned}\right.
\end{equation}
and it can used for both dilute and dense flow.
$C_d$ is the drag coefficient, which is obtained by,
\begin{equation}
C_d = \left\{\begin{aligned}
&\frac{24}{Re_s}\left(1+0.15 Re_s^{0.687}\right), &  & Re_s \le 1000, \\
&0.44, &  & Re_s > 1000,
\end{aligned} \right.
\end{equation}
where $d_s$ is the diameter of solid particle, and $\mu_g$ is the dynamic viscosity of gas phase. $Re_s = |\textbf{U}_g-\textbf{u}| d_s/\nu_g$ is the particle Reynolds number, and $\nu_g=\mu_g/\rho_g$ is the kinematic viscosity of gas phase. Besides, another interactive force considered is the buoyancy force, which can be modeled as,
\begin{gather}\label{buoyancy force model}
\textbf{F}_b = -\frac{m_s}{\rho_{s}} \nabla_x p_g,
\end{gather}
where $p_g$ is the pressure of gas phase. Then, the acceleration term can be obtained,
\begin{gather*}\label{particle phase acceleration term}
\textbf{a}=\frac{\textbf{D} + \textbf{F}_b}{m_s} + \textbf{G}.
\end{gather*}

When the collision between solid particles are elastic with $r=1$, in the continuum flow regime the hydrodynamic equations becomes the Euler equations which can be obtained based on the Chapman-Enskog asymptotic analysis,
\begin{align}\label{particle phase Euler equ}
&\frac{\partial \left(\epsilon_s\rho_s\right)}{\partial t}
+ \nabla_x \cdot \left(\epsilon_s\rho_s \textbf{U}_s\right) = 0,\nonumber \\
&\frac{\partial \left(\epsilon_s\rho_s \textbf{U}_s\right)}{\partial t}
+ \nabla_x \cdot \left(\epsilon_s\rho_s \textbf{U}_s \textbf{U}_s + p_s \mathbb{I} \right)
= \frac{\epsilon_{s}\rho_{s}\left(\textbf{U}_g - \textbf{U}_s\right)}{\tau_{st}}
- \epsilon_{s} \nabla_x p_g
+ \epsilon_{s}\rho_{s} \textbf{G} , \\
&\frac{\partial \left(\epsilon_s\rho_s E_s\right)}{\partial t}
+ \nabla_x \cdot \left(\left(\epsilon_s\rho_s E_s  + p_s\right) \textbf{U}_s \right)
= \frac{\epsilon_{s}\rho_{s}\textbf{U}_s \cdot \left(\textbf{U}_g - \textbf{U}_s\right)}{\tau_{st}}
- \frac{3p_s}{\tau_{st}}
- \epsilon_{s} \textbf{U}_s \cdot \nabla_x p_g
+ \epsilon_{s}\rho_{s} \textbf{U}_s \cdot \textbf{G}.\nonumber
\end{align}
Note that the heat conduction between the particle and gas phase are neglected in this paper. In summary, the evolution of particle phase is governed by Eq.\eqref{particle phase kinetic equ}, and the hydrodynamic equations Eq.\eqref{particle phase Euler equ} is only the asymptotic solution in the continuum flow limit for the solid-particle phase.

\subsection{UGKWP method}
In this subsection, the UGKWP for the evolution of particle phase is introduced. Generally, the kinetic equation of particle phase Eq.\eqref{particle phase kinetic equ} is split as,
\begin{align}
\label{particle phase kinetic equ without acce}
\mathcal{L}_{s1} &:~~ \frac{\partial f_{s}}{\partial t}
+ \nabla_x \cdot \left(\textbf{u}f_{s}\right)
= \frac{g_{s}-f_{s}}{\tau_{s}}, \\
\label{particle phase kenetic equ only acce}
\mathcal{L}_{s2} &:~~ \frac{\partial f_{s}}{\partial t}
+ \nabla_u \cdot \left(\textbf{a}f_{s}\right)
= 0,
\end{align}
and splitting operator is used to solve Eq.\eqref{particle phase kinetic equ}. Firstly we focus on $\mathcal{L}_{s1}$ part, the particle phase kinetic equation without external force,
\begin{gather*}
\frac{\partial f_{s}}{\partial t}
+ \nabla_x \cdot \left(\textbf{u}f_{s}\right)
= \frac{g_{s}-f_{s}}{\tau_{s}}.
\end{gather*}
For brevity, the subscript $s$ standing for the solid particle phase will be neglected in this subsection. The integration solution of the kinetic equation can be written as,
\begin{equation}\label{particle phase integration solution}
f(\textbf{x},t,\textbf{u})=\frac{1}{\tau}\int_0^t g(\textbf{x}',t',\textbf{u} )e^{-(t-t')/\tau}\text{d}t'\\
+e^{-t/\tau}f_0(\textbf{x}-\textbf{u}t, \textbf{u}),
\end{equation}
where $\textbf{x}'=\textbf{x}+\textbf{u}(t'-t)$ is the trajectory of particles, $f_0$ is the initial gas distribution function at time $t=0$, and $g$ is the corresponding equilibrium state.

In UGKWP, both macroscopic conservative variables and microscopic gas distribution function need to be updated. Generally, in the finite volume framework, the cell-averaged macroscopic variables $\textbf{W}_i$ of cell $i$ can be updated by the conservation law,
\begin{gather}
\textbf{W}_i^{n+1} = \textbf{W}_i^n - \frac{1}{\Omega_i} \sum_{S_{ij}\in \partial \Omega_i}\textbf{F}_{ij}S_{ij} + \Delta t \textbf{S}_{i},
\end{gather}
where $\textbf{W}_i=\left(\rho_i, \rho_i \textbf{U}_i, \rho_i E_i\right)$ is the cell-averaged macroscopic variables,
\begin{gather*}
\textbf{W}_i = \frac{1}{\Omega_{i}}\int_{\Omega_{i}} \textbf{W}\left(\textbf{x}\right) \text{d}\Omega,
\end{gather*}
$\Omega_i$ is the volume of cell $i$, $\partial\Omega_i$ denotes the set of cell interfaces of cell $i$, $S_{ij}$ is the area of the $j$-th interface of cell $i$, $\textbf{F}_{ij}$ denotes the macroscopic fluxes across the interface $S_{ij}$, which can be written as
\begin{align}\label{particle phase Flux equation}
\textbf{F}_{ij}=\int_{0}^{\Delta t} \int \textbf{u}\cdot\textbf{n}_{ij} f_{ij}(\textbf{x},t,\textbf{u}) \bm{\psi} \text{d}\textbf{u}\text{d}t,
\end{align}
where $\textbf{n}_{ij}$ denotes the normal vector of interface $S_{ij}$, $f_{ij}\left(t\right)$ is the time-dependent distribution function on the interface $S_{ij}$, and $\bm{\psi}=(1,\textbf{u},\displaystyle \frac{1}{2}\textbf{u}^2)^T$. $\textbf{S}_{i}$ is the source term due to inelastic collision inside each control volume, where the solid-particle's internal energy has not been taken into account in the above equation.

Substituting the time-dependent distribution function Eq.\eqref{particle phase integration solution} into Eq.\eqref{particle phase Flux equation}, the fluxes can be obtained,
\begin{align*}
\textbf{F}_{ij}
&=\int_{0}^{\Delta t} \int \textbf{u}\cdot\textbf{n}_{ij} f_{ij}(\textbf{x},t,\textbf{u}) \bm{\psi} \text{d}\textbf{u}\text{d}t\\
&=\int_{0}^{\Delta t} \int\textbf{u}\cdot\textbf{n}_{ij} \left[ \frac{1}{\tau}\int_0^t g(\textbf{x}',t',\textbf{u})e^{-(t-t')/\tau}\text{d}t' \right] \bm{\psi} \text{d}\textbf{u}\text{d}t\\
&+\int_{0}^{\Delta t} \int\textbf{u}\cdot\textbf{n}_{ij} \left[ e^{-t/\tau}f_0(\textbf{x}-\textbf{u}t,\textbf{u})\right] \bm{\psi} \text{d}\textbf{u}\text{d}t\\
&\overset{def}{=}\textbf{F}^{eq}_{ij} + \textbf{F}^{fr}_{ij}.
\end{align*}

The procedure of obtaining the local equilibrium state $g_0$ at the cell interface as well as the construction of $g\left(t\right)$ is the same as that in GKS.
For a second-order accuracy, the equilibrium state $g$ around the cell interface is written as,
\begin{gather*}
g\left(\textbf{x}',t',\textbf{u}\right)=g_0\left(\textbf{x},\textbf{u}\right)
\left(1 + \overline{\textbf{a}} \cdot \textbf{u}\left(t'-t\right) + \bar{A}t'\right),
\end{gather*}
where $\overline{\textbf{a}}=\left[\overline{a_1}, \overline{a_2}, \overline{a_3}\right]^T$, $\overline{a_i}=\frac{\partial g}{\partial x_i}/g$, $i=1,2,3$,  $\overline{A}=\frac{\partial g}{\partial t}/g$, and $g_0$ is the local equilibrium on the interface.
Specifically, the coefficients of spatial derivatives $\overline{a_i}$ can be obtained from the corresponding derivatives of the macroscopic variables,
\begin{equation*}
\left\langle \overline{a_i}\right\rangle=\partial \textbf{W}_0/\partial x_i,
\end{equation*}
where $i=1,2,3$, and $\left\langle...\right\rangle$ means the moments of the Maxwellian distribution functions,
\begin{align*}
\left\langle...\right\rangle=\int \bm{\psi}\left(...\right)g\text{d}\textbf{u}.
\end{align*}
The coefficients of temporal derivative $\overline{A}$ can be determined by the compatibility condition,
\begin{equation*}
\left\langle \overline{\textbf{a}} \cdot \textbf{u}+\overline{A} \right\rangle =
\left[\begin{array}{c}
0\\
\textbf{0}\\
-\frac{Q_{loss}}{\tau_s}
\end{array}\right].
\end{equation*}
where $Q_{loss}=\frac{\left(1-r^2\right)3p_s}{2}$, caused by the particle-particle inelastic collision. Now, all the coefficients in the equilibrium state $g\left(\textbf{x}',t',\textbf{u}\right)$ have been determined, and its integration becomes,
\begin{gather}
f^{eq}(\textbf{x},t,\textbf{u}) \overset{def}{=} \frac{1}{\tau}\int_0^t g(\textbf{x}',t',\textbf{u})e^{-(t-t')/\tau}\text{d}t' \nonumber\\
= c_1 g_0\left(\textbf{x},\textbf{u}\right)
+ c_2 \overline{\textbf{a}} \cdot \textbf{u} g_0\left(\textbf{x},\textbf{u}\right)
+ c_3 A g_0\left(\textbf{x},\textbf{u}\right),
\end{gather}
with coefficients,
\begin{align*}
c_1 &= 1-e^{-t/\tau}, \\
c_2 &= \left(t+\tau\right)e^{-t/\tau}-\tau, \\
c_3 &= t-\tau+\tau e^{-t/\tau},
\end{align*}
and thereby the integrated flux over a time step for equilibrium state can be obtained,
\begin{gather*}
\textbf{F}^{eq}_{ij}
=\int_{0}^{\Delta t} \int \textbf{u}\cdot\textbf{n}_{ij} f_{ij}^{eq}(\textbf{x},t,\textbf{u})\bm{\psi}\text{d}\textbf{u}\text{d}t.
\end{gather*}

Besides, the flux contribution from the particle free transport $f_0$ is calculated by tracking the particles sampled from $f_0$. Therefore, the updating of the cell-averaged macroscopic variables can be written as,
\begin{gather}\label{particle phase equ_updateW_ugkp}
\textbf{W}_i^{n+1} = \textbf{W}_i^n - \frac{1}{\Omega_i} \sum_{S_{ij}\in \partial \Omega_i}\textbf{F}^{eq}_{ij}S_{ij}
+ \frac{\textbf{w}_{i}^{fr}}{\Omega_{i}}
+ \Delta t \textbf{S}_{i},
\end{gather}
where $\textbf{w}^{fr}_i$ is the net free streaming flow of cell $i$, standing for the flux contribution of the free streaming of particles, and the term $\textbf{S}_{i} = \left[0,\textbf{0},-\frac{Q_{loss}}{\tau_s}\right]^T$ is the source term due to the inelastic collision for solid particle phase.

The net free streaming flow $\textbf{w}^{fr}_i$ is determined in the following. The evolution of particle should also satisfy the integral solution of the kinetic equation, which can be written as,
\begin{equation}
f(\textbf{x},t,\textbf{u})
=\left(1-e^{-t/\tau}\right)g^{+}(\textbf{x},t,\textbf{u})
+e^{-t/\tau}f_0(\textbf{x}-\textbf{u}t,\textbf{u}),
\end{equation}
where $g^{+}$ is named as the hydrodynamic distribution function with analytical formulation. The initial distribution function $f_0$ have a probability of $e^{-t/\tau}$ to free transport and $1-e^{-t/\tau}$ to colliding with other particles. The post-collision particles satisfies the distribution $g^+\left(\textbf{x},\textbf{u},t\right)$. The free transport time before the first collision with other particles is denoted as $t_c$. The cumulative distribution function of $t_c$ is,
\begin{gather}\label{particle phase wp cumulative distribution}
F\left(t_c < t\right) = 1 - e^{-t/ \tau},
\end{gather}
and therefore $t_c$ can be sampled as $t_c=-\tau \text{ln}\left(\eta\right)$, where $\eta$ is a random number generated from a uniform distribution $U\left(0,1\right)$. Then, the free streaming time $t_f$ for particle $k$ is determined by,
\begin{gather}
t_f = min\left[-\tau\text{ln}\left(\eta\right), \Delta t\right],
\end{gather}
where $\Delta t$ is the time step. Therefore, within one time step, all particles can be divided into two groups: the collisionless particle and the collisional particle, and they are determined by the relation between of time step $\Delta t$ and free streaming time $t_f$. Specifically, if $t_f=\Delta t$ for one particle, it is collisionless particle, and the trajectory of this particle is fully tracked in the whole time step. On the contrary, if $t_f<\Delta t$ for one particle, it is collisional particle, and its trajectory will be tracked until $t_f$. The collisional particle is eliminated at $t_f$ in the simulation and the associated mass, momentum and energy carried by this particle are merged into the macroscopic quantities in the relevant cell by counting its contribution through the flux function. More specifically, the particle trajectory in the free streaming process within time $t_f$ is tacked by,
\begin{gather}
\textbf{x} = \textbf{x}^n + \textbf{u}^n t_f .
\end{gather}
The term $\textbf{w}_{i}^{fr}$ can be calculated by counting the particles passing through the interfaces of cell $i$,
\begin{gather}
\textbf{w}_{i}^{fr} = \sum_{k\in P\left(\partial \Omega_{i}^{+}\right)} \bm{\phi}_k - \sum_{k\in P\left(\partial \Omega_{i}^{-}\right)} \bm{\phi}_k,
\end{gather}
where, $P\left(\partial \Omega_{i}^{+}\right)$ is the particle set moving into the cell $i$ during one time step, $P\left(\partial \Omega_{i}^{-}\right)$ is the particle set moving out of the cell $i$ during one time step, $k$ is the particle index in one specific set, and $\bm{\phi}_k=\left[m_{k}, m_{k}\textbf{u}_k, \frac{1}{2}m_{k}(\textbf{u}^2_k)\right]^T$ is the mass, momentum and energy carried by particle $k$. Therefore, $\textbf{w}_{i}^{fr}/\Omega_{i}$ is the net conservative quantities caused by the free stream of the tracked particles. Now, all the terms in Eq.\eqref{particle phase equ_updateW_ugkp} have been determined and the macroscopic variables $\textbf{W}_i$ can be updated.

The trajectories of all particles have been tracked during the time interval $\left(0, t_f\right)$. For the collisionless particles with $t_f=\Delta t$, they still survive at the end of one time step; while the collisional particles with $t_f<\Delta t$ are deleted after their first collision and they are supposed to go to the equilibrium state in that cell. Therefore, the macroscopic variables of the collisional particles in cell $i$ at the end of each time step can be directly obtained based on the conservation law,
\begin{gather}
\textbf{W}^h_i = \textbf{W}^{n+1}_i - \textbf{W}^p_i,
\end{gather}
where $\textbf{W}^{n+1}_i$ is the updated conservative variables in Eq.\eqref{particle phase equ_updateW_ugkp} and  $\textbf{W}^p_i$ are the  mass, momentum, and energy of remaining collisionless particles in the cell at the end of the time step.
Besides, the macroscopic variables $\textbf{W}^h_i$ account for all eliminated collisional particles to the equilibrium state,
and these particles can be re-sampling from $\textbf{W}^h_i$ based on the overall Maxwellian distribution at the beginning of next time step.
Now the updates of both macroscopic variables and the microscopic particles have been presented. The above method is the so-called unified gas-kinetic particle (UGKP) method.

The above UGKP can be further developed to UGKWP method.
In UGKP method, all particles are divided into collisionless and collisional particles in each time step. The collisional particles are deleted after the first collision and re-sampled from $\textbf{W}^h_i$ at the beginning of next time step.
However, only the collisionless part of the re-samples particles can survive in the next time step, and all collisional ones will be deleted again.
Actually, the transport fluxes from these collisional particles can be evaluated analytically without using particles.
According to the cumulative distribution Eq.\eqref{particle phase wp cumulative distribution}, the proportion of the collisionless particles is $e^{-\Delta t/\tau}$, and therefore in UGKWP only the collisionless particles from the hydrodynamic variables $\textbf{W}^{h}_i$ in cell $i$ will  be re-sampled with the total mass, momentum, and energy,
\begin{gather}
\textbf{W}^{hp}_i = e^{-\Delta t/\tau} \textbf{W}^{h}_i.
\end{gather}
Then, the free transport time of all the re-sampled particles will be $t_f=\Delta t$ in UGKWP.
The fluxes $\textbf{F}^{fr,wave}$ from these un-sampled collisional particle of $ (1- e^{-\Delta t/\tau} )\textbf{W}^{h}_i$ can be evaluated
analytically \cite{WP-first-liu2020unified, WP-second-zhu-unstructured-mesh-zhu2019unified}.
Now, same as UGKP, the net flux $\textbf{w}_{i}^{fr,p}$ by the free streaming of the existing particles in UGKWP can be calculated by
\begin{gather}
\textbf{w}_{i}^{fr,p} = \sum_{k\in P\left(\partial \Omega_{i}^{+}\right)} \bm{\phi}_k - \sum_{k\in P\left(\partial \Omega_{i}^{-}\right)} \bm{\phi}_k.
\end{gather}
Then, the macroscopic flow variables in UGKWP are updated by
\begin{gather}\label{particle phase wp final update W}
\textbf{W}_i^{n+1} = \textbf{W}_i^n
- \frac{1}{\Omega_i} \sum_{S_{ij}\in \partial \Omega_i}\textbf{F}^{eq}_{ij}S_{ij}
- \frac{1}{\Omega_i} \sum_{S_{ij}\in \partial \Omega_i}\textbf{F}^{fr,wave}_{ij}S_{ij}
+ \frac{\textbf{w}_{i}^{fr,p}}{\Omega_{i}}
+ \Delta t \textbf{S}_{i}.
\end{gather}

The second part $\mathcal{L}_{s2}$ in Eq.\eqref{particle phase kenetic equ only acce} accounts for the external acceleration,
\begin{gather*}
\frac{\partial f_{s}}{\partial t}
+ \nabla_u \cdot \left(\textbf{a}f_{s}\right)
= 0,
\end{gather*}
where the velocity-dependent acceleration term caused by inter-phase forces and solid particle's gravity has the following form,
\begin{gather*}
\textbf{a} = \frac{\textbf{U}_g - \textbf{u}}{\tau_{st}} - \frac{1}{\rho_{s}} \nabla_x p_g + \textbf{G}.
\end{gather*}
Taking moment $\bm{\psi}$ to Eq.\eqref{particle phase kenetic equ only acce},
\begin{gather*}
\int \bm{\psi}
\left( \frac{\partial f_{s}}{\partial t}
+ \textbf{a} \cdot \nabla_u f_{s}
+ f_{s}\nabla_u \cdot \textbf{a}
\right) \text{d}\textbf{u} = 0,
\end{gather*}
and in the Euler regime with $f_s = g_s + \mathcal{O}\left(\tau_{s}\right)$, we can obtain,
\begin{gather*}
\frac{\partial \textbf{W}_s}{\partial t} + \textbf{Q}_s= 0,
\end{gather*}
where
\begin{gather*}
\textbf{W}_s=\left[\begin{array}{c}
\epsilon_s\rho_s\\
\epsilon_s\rho_s \textbf{U}_s\\
\epsilon_s\rho_s E_s
\end{array}
\right], ~~
\textbf{Q}_s=\left[\begin{array}{c}
0 \\
\frac{\epsilon_s\rho_s\left(\textbf{U}_s-\textbf{U}_g\right)}{\tau_{st}}
+\epsilon_s \nabla_x p_g
- \epsilon_{s}\rho_{s} \textbf{G} \\
\frac{\epsilon_s\rho_{s}\textbf{U}_s \cdot \left(\textbf{U}_s-\textbf{U}_g\right)}{\tau_{st}} +3\frac{p_s}{\tau_{st}}
+ \epsilon_s\textbf{U}_s \cdot \nabla_x p_g
- \epsilon_{s}\rho_{s} \textbf{U}_s \cdot \textbf{G}
\end{array}\right].
\end{gather*}
When the first-order forward Euler method is employed for time marching, the cell-averaged macroscopic variable can be updated by,
\begin{gather}\label{update macroscopic variable of acceleration wave}
\textbf{W}^{n+1}_s = \textbf{W}_s - \textbf{Q}_s \Delta t,
\end{gather}
and the modifications on velocity and location of the remaining free transport particles can be written as,
\begin{align}
\textbf{u}^{n+1} &= \textbf{u} + \textbf{a}t_f,\\
\textbf{x}^{n+1} &= \textbf{x} + \frac{\textbf{a}}{2} t_f^2.\label{displacement by acceleartion term}
\end{align}
Now the update of the particle phase in one time step has been finished. In the following, specific variables determination for the solid-particle
phase will be presented.

\subsection{Particle phase Knudsen number}
The particle phase regime is determined by its Knudsen number $Kn$, defined by the ratio of collision time of solid particles $\tau_{s}$ to the characteristic time of macroscopic flow $t_{ref}$,
\begin{gather}\label{particle phase Kn_s}
Kn = \frac{\tau_s}{t_{ref}}.
\end{gather}
Specifically, $\tau_s$ is the time interval between collisions of solid particles, or called the particle collision time, and $t_{ref}$ is the characteristic time, defined as the ratio flow characteristic length to the flow characteristic velocity, $t_{ref}=L_{ref}/U_{ref}$.
According to the previous studies \cite{Gasparticle-MOM-Fox-passalacqua2010fully, Gasparticle-momentmethod-Fox2013computational}, in this paper $\tau_s$ is taken as,
\begin{gather}\label{particle phase tau_s}
\tau_s = \frac{\sqrt{\pi}d_s}{12\epsilon_sg_0}\sqrt{2\lambda_s},
\end{gather}
where $d_s$ is the diameter of solid particle, $\epsilon_s$ is the volume fraction of solid phase. $g_0$ is the radial distribution function with the following form,
\begin{gather}
g_0 = \frac{2-c}{2\left(1-c\right)^3},
\end{gather}
where $c=\epsilon_s/\epsilon_{s,max}$ is the ratio of the volume fraction $\epsilon_{s}$ to the allowed maximum value $\epsilon_{s,max}$. Generally, for dilute particulate flow, $\tau_{s}$ is more likely much larger than $t_{ref}$, leading to a large $Kn$, and the particle transport plays more important role in the evolution. However, for dense particulate flow, the collision between solid particles is in high-frequency, which results in a small $\tau_{s}$ and thereby a small $Kn$, and the inter-particle collision plays the key effect in the evolution.

\subsection{Frictional pressure}
When the solid phase is in high concentration, the frictional pressure $p_{fric}$ has to be considered. $p_{fric}$ accounts for the enduring inter-particle contacts and frictions, which plays important roles in the near packing situation. Some expressions for $p_{fric}$ have been proposed in the previous studies \cite{Gasparticle-KTGF-pressure-friction-johnson1987frictional, Gasparticle-pressure-friction-srivastava2003analysis, Gasparticle-pressure-friction-schneiderbauer2012comprehensive}. In this paper, the correlation proposed by Johnson and Jackson is employed \cite{Gasparticle-KTGF-pressure-friction-johnson1987frictional}, which can be written as,
\begin{align}
p_{fric} = \left\{\begin{aligned}
&~~~~~~~~ 0 &  ,   & \epsilon_{s} \le \epsilon_{s,crit}, \\
&0.1 \epsilon_{s} \frac{\left(\epsilon_{s} - \epsilon_{s,crit}\right)^2}{\left(\epsilon_{s,max} -  \epsilon_{s}\right)^5}&  ,   & \epsilon_{s} > \epsilon_{s,crit},
\end{aligned} \right.
\end{align}
where $\epsilon_{s,crit}$ is the critical volume fraction of particle flow, and it takes a value $0.5$ in this paper unless special notification. Therefore, the momentum and energy equation in Eq.\eqref{particle phase Euler equ} will be rewritten as,
\begin{gather}\label{particle phase momentum equ equ with p_fr}
\frac{\partial \left(\epsilon_s\rho_s \textbf{U}_s\right)}{\partial t}
+ \nabla_x \cdot \left(\epsilon_s\rho_s \textbf{U}_s \textbf{U}_s + p_s \mathbb{I} + p_{fric} \mathbb{I}\right)
= \frac{\epsilon_{s}\rho_{s}\left(\textbf{U}_g - \textbf{U}_s\right)}{\tau_{st}}
- \epsilon_{s} \nabla_x p_g
+ \epsilon_{s}\rho_{s} \textbf{G}.
\end{gather}
\begin{gather}\label{particle phase energy equ equ with p_fr}
\frac{\partial \left(\epsilon_s\rho_s E_s\right)}{\partial t}
+ \nabla_x \cdot \left(\left(\epsilon_s\rho_s E_s  + p_s + p_{fric}\right) \textbf{U}_s \right)
= \frac{\epsilon_{s}\rho_{s}\textbf{U}_s \cdot \left(\textbf{U}_g - \textbf{U}_s\right)}{\tau_{st}}
- \frac{3p_s}{\tau_{st}}
- \epsilon_{s} \textbf{U}_s \cdot \nabla_x p_g \nonumber\\
~~~~~~~~~~~~~~~~~~~~~~~~~~
+ \epsilon_{s}\rho_{s} \textbf{U}_s \cdot \textbf{G}.
\end{gather}
The terms relevant to frictional pressure, $\nabla_x \cdot \left(p_{fric} \mathbb{I}\right)$ and $\nabla_x \cdot \left(p_{fric} \textbf{U}_s\right)$, are solved as source terms in this paper.

\subsection{Flux limiting model near the packing condition}
The introduction of frictional pressure $p_{fric}$ can avoid the solid particles' over-assembling since it increases dramatically when the particle phase approaches its limiting packing state \cite{Gasparticle-KTGF-pressure-friction-johnson1987frictional, Gasparticle-TFM-compressible-houim2016multiphase}. Besides, a flux limiting model is proposed in this paper to effectively prevent the solid volume fraction $\epsilon_{s}$ from exceeding its maximum value $\epsilon_{s,max}$. Taking one-dimensional example, in UGKWP the numerical flux at interface $i+1/2$ between cell $i$ and cell $i+1$ can be generally written as,
\begin{equation}\label{particle phase flux left rigth form original}
\textbf{F}_{i+1/2}
= \int_{0}^{\Delta t} \int_{u>0} u f_{i+1/2}(\textbf{x},t) \bm{\psi} \text{d}u\text{d}t
+ \int_{0}^{\Delta t} \int_{u<0} u f_{i+1/2}(\textbf{x},t) \bm{\psi} \text{d}u\text{d}t\\
\overset{def}{=} \textbf{F}_{i+1/2}^{l} + \textbf{F}_{i+1/2}^{r},
\end{equation}
which will be modified as,
\begin{equation}\label{particle phase flux left right form flux limiting}
\textbf{F}_{i+1/2} =
\textbf{C}\left[\alpha\left(\epsilon_{s,i+1}\right)\right] \cdot \textbf{F}_{i+1/2}^{l}
+ \textbf{C}\left[\alpha\left(\epsilon_{s,i}\right)\right] \cdot \textbf{F}_{i+1/2}^{r},
\end{equation}
with
\begin{equation*}
\textbf{C}\left[\alpha\right] =
\begin{bmatrix}
1-\alpha & 0 & 0 \\
0 & 1+\alpha & 0 \\
0 & 0 & 1-\alpha
\end{bmatrix},
\end{equation*}
where $\alpha$ is the limiting factor, and it depends on the cell-averaged solid volume fraction $\epsilon_{s}$ as,
\begin{equation}\label{packing limit factor}
\alpha\left(\epsilon_{s}\right) = \left\{\begin{aligned}
&~~~~~~~~ 0 &  ,   & \epsilon_{s} \le k\epsilon_{s,max}, \\
&\left(\frac{\epsilon_{s} - k\epsilon_{s,max}}{\epsilon_{s,max} - k \epsilon_{s,max}}\right)^2&  ,   & \epsilon_{s} > k\epsilon_{s,max}.
\end{aligned} \right.
\end{equation}
Here $k$ is a threshold for the flux limiting model, and it takes a value $0.95$ unless special notification in this paper. As shown in Eq.\eqref{packing limit factor}, when $\epsilon_{s}$ is smaller than $k\epsilon_{s,max}$, the limiting factor $\alpha$ goes to $0$ and there is no limiting; while when $\epsilon_{s}$ is larger than $k\epsilon_{s,max}$, $\alpha$ will increase and the limiting model works. Particularly, when the packing limit approaches to $\epsilon_{s}=\epsilon_{s,max}$, $\alpha$ also takes its maximum value $1$. As a result, solid particles cannot flow into the ``saturated" cell, and the solid volume fraction $\epsilon_{s}$ will not increase anymore.
In addition, Eq.\eqref{particle phase flux left right form flux limiting} indicates that, as this limiting model is activated, only the "inflow" across the interface will be effected, while the "outflow" will not be limited as a physical modeling to the reality.

\section{GKS for gas phase}
\subsection{Governing equation for gas phase}
The gas phase is regarded as continuum flow and the governing equations are the Navier-Stokes equations with source terms reflecting the inter-phase interaction \cite{Gasparticle-book-gidaspow1994multiphase, Gasparticle-book-ishii2010thermo},
\begin{align}\label{gas phase macroscopic equ}
&\frac{\partial \left(\widetilde{\rho_g}\right)}{\partial t}
+ \nabla_x \cdot \left(\widetilde{\rho_g} \textbf{U}_g\right)= 0,\nonumber \\
&\frac{\partial \left(\widetilde{\rho_g} \textbf{U}_g\right)}{\partial t}
+ \nabla_x \cdot \left(\widetilde{\rho_g} \textbf{U}_g \textbf{U}_g + \widetilde{p_g}\mathbb{I}\right)
- \epsilon_{g} \nabla_x \cdot \left(\mu_g \bm{\sigma}\right)
=
p_g \nabla_x \epsilon_{g}
-\frac{\epsilon_{s}\rho_{s}\left(\textbf{U}_g - \textbf{U}_s\right)}{\tau_{st}}
+ \widetilde{\rho_g} \textbf{G} , \\
&\frac{\partial \left(\widetilde{\rho_g} E_g\right)}{\partial t}
+ \nabla_x \cdot \left(\left(\widetilde{\rho_g} E_g  + \widetilde{p_g}\right) \textbf{U}_g \right)
- \epsilon_{g} \nabla_x \cdot \left(\mu_g \bm{\sigma}\cdot\textbf{U}_g - \kappa \nabla_x T_g \right)
=
- p_{g} \frac{\partial \epsilon_{g}}{\partial t} \nonumber \\
& ~~~~~~~~~~~~~~~~~~~~~~~~~~~~~~~~~~~~~~~~~~~~~~~~~~
-\frac{\epsilon_{s}\rho_{s}\textbf{U}_s \cdot \left(\textbf{U}_g - \textbf{U}_s\right)}{\tau_{st}}
+ \frac{3p_s}{\tau_{st}}
+ \widetilde{\rho_g} \textbf{U}_g \cdot \textbf{G}, \nonumber
\end{align}
where $\widetilde{\rho_g}=\epsilon_{g}\rho_g$ is the apparent density of gas phase, $p_g=\rho_gRT_g$ is the pressure of gas phase and $\widetilde{p_g}=\widetilde{\rho_g}RT_g$, the strain rate tensor $\bm{\sigma}$ is
\begin{gather*}
\bm{\sigma} = \nabla_x\textbf{U}_g + \left(\nabla_x\textbf{U}_g\right)^T
- \frac{2}{3} \nabla_x \cdot \textbf{U}_g \mathbb{I},
\end{gather*}
and
\begin{gather*}
\mu_g = \tau_{g} p_g, ~~~~ \kappa = \frac{5}{2} R \tau_{g} p_g.
\end{gather*}
In particular, at the right hand side in Eq.\eqref{gas phase macroscopic equ}, the term $p_{g} \nabla_x \epsilon_{g}$ is called ``nozzle" term, and the associated work term $- p_{g} \frac{\partial \epsilon_{g}}{\partial t}$ is called $pDV$ work term, since it is similar to the $pDV$ term in the quasi-one-dimensional gas nozzle flow equations \cite{Gasparticle-TFM-compressible-houim2016multiphase}. Unphysical pressure fluctuations might occurs if the ``nozzle" term and $pDV$ term are not solved correctly. According to \cite{Toro2013book}, Eq.\eqref{gas phase macroscopic equ} can be written as the following form,
\begin{align}\label{gas phase macroscopic equ final}
&\frac{\partial \left(\rho_g\right)}{\partial t}
+ \nabla_x \cdot \left(\rho_g \textbf{U}_g\right)= C_{\epsilon_g}\rho_g,\nonumber \\
&\frac{\partial \left(\rho_g \textbf{U}_g\right)}{\partial t}
+ \nabla_x \cdot \left(\rho_g \textbf{U}_g \textbf{U}_g + p_g\mathbb{I} - \mu_g \bm{\sigma}\right)
=
C_{\epsilon_g} \rho_g \textbf{U}_g
-\frac{\epsilon_{s}\rho_{s}\left(\textbf{U}_g - \textbf{U}_s\right)}{\epsilon_g \tau_{st}}
+ \rho_g\textbf{G} , \\
&\frac{\partial \left(\rho_g E_g\right)}{\partial t}
+ \nabla_x \cdot \left(\left(\rho_g E_g  + p_g\right) \textbf{U}_g
- \mu_g \bm{\sigma}\cdot\textbf{U}_g + \kappa \nabla_x T_g \right)
=
C_{\epsilon_g} \left(\rho_g E_g + p_g\right) \nonumber \\
& ~~~~~~~~~~~~~~~~~~~~~~~~~~~~~~~~~~~~~~~~~~
-\frac{\epsilon_{s}\rho_{s}\textbf{U}_s \cdot \left(\textbf{U}_g - \textbf{U}_s\right)}{\epsilon_g \tau_{st}}
+ \frac{3p_s}{\epsilon_g \tau_{st}}
+ \rho_g \textbf{U}_g \cdot \textbf{G}, \nonumber
\end{align}
where, $C_{\epsilon_g} = -\frac{1}{\epsilon_{g}}\frac{\text{d}\epsilon_{g}}{\text{d}t}$ with $\frac{\text{d}\epsilon_{g}}{\text{d}t}=\frac{\partial \epsilon_{g}}{\partial t}+\textbf{U}_g \cdot \nabla\epsilon_{g}$, and how to solve $C_{\epsilon_{g}}$ in this paper will be introduced later.

\subsection{GKS for gas evolution}
This subsection introduces the evolution of gas phase in gas-particle two-phase system. The gas flow is governed by the Navier-Stokes equations with the inter-phase interaction, and the corresponding  GKS is a limiting scheme of UGKWP in the continuum flow regime. In general, the evolution of gas phase Eq.\eqref{gas phase macroscopic equ final} can be split into two parts,

\begin{align}
\mathcal{L}_{g1}&:~~
\left\{
\begin{array}{lr}
\frac{\partial \left(\rho_g\right)}{\partial t}
+ \nabla_x \cdot \left(\rho_g \textbf{U}_g\right)= 0, & \vspace{1ex}\\
\frac{\partial \left(\rho_g \textbf{U}_g\right)}{\partial t}
+ \nabla_x \cdot \left(\rho_g \textbf{U}_g \textbf{U}_g + p_g\mathbb{I} - \mu_g \bm{\sigma}\right)
= 0, & \vspace{1ex}\\
\frac{\partial \left(\rho_g E_g\right)}{\partial t}
+ \nabla_x \cdot \left(\left(\rho_g E_g  + p_g\right) \textbf{U}_g
- \mu_g \bm{\sigma}\cdot\textbf{U}_g + \kappa \nabla_x T_g \right) = 0, &
\end{array}
\right. \\
\nonumber\\
\mathcal{L}_{g2}&:~~
\left\{
\begin{array}{lr}
\frac{\partial \left(\rho_g\right)}{\partial t} = C_{\epsilon_g}\rho_g, & \vspace{1ex}\\
\frac{\partial \left(\rho_g \textbf{U}_g\right)}{\partial t} =
C_{\epsilon_g} \rho_g \textbf{U}_g
-\frac{\epsilon_{s}\rho_{s}\left(\textbf{U}_g - \textbf{U}_s\right)}{\epsilon_g \tau_{st}}
+ \rho_g\textbf{G}, & \vspace{1ex}\\
\frac{\partial \left(\rho_g E_g\right)}{\partial t} =
C_{\epsilon_g} \left(\rho_g E_g + p_g\right)
-\frac{\epsilon_{s}\rho_{s}\textbf{U}_s \cdot \left(\textbf{U}_g - \textbf{U}_s\right)}{\epsilon_g \tau_{st}}
+ \frac{3p_s}{\epsilon_g \tau_{st}}
+ \rho_g \textbf{U}_g \cdot \textbf{G}. &
\end{array}
\right.
\end{align}
The GKS is constructed to solve $\mathcal{L}_{g1}$ and $\mathcal{L}_{g2}$ separately.
Firstly, the kinetic equation without acceleration term for gas phase $\mathcal{L}_{g1}$ is,
\begin{equation}\label{gas phase kinetic equ without acce}
\frac{\partial f_{g}}{\partial t}
+ \nabla_x \cdot \left(\textbf{u}f_{g}\right)
= \frac{g_{g}-f_{g}}{\tau_{g}},
\end{equation}
where $\textbf{u}$ is the velocity, $\tau_g$ is the relaxation time for gas phase, $f_{g}$ is the distribution function of gas phase, and $g_{g}$ is the corresponding equilibrium state (Maxwellian distribution). The local equilibrium state $g_{g}$ can be written as,
\begin{gather*}
g_{g}=\rho_g\left(\frac{\lambda_g}{\pi}\right)^{\frac{K+3}{2}}e^{-\lambda_g\left[(\textbf{u}-\textbf{U}_g)^2+\bm{\xi}^2\right]},
\end{gather*}
where $\rho_g$ is the density of gas phase. $\lambda_g$ is determined by gas temperature through $\lambda_g = \frac{m_g}{2k_BT_g}$, where $m_g$ is the molecular mass, $\textbf{U}_g$ is the macroscopic velocity of gas phase. $K$ is the internal degree of freedom with $K=(5-3\gamma)/(\gamma-1)$ for three-dimensional diatomic gas, where $\gamma=1.4$ is the specific heat ratio.
The collision term satisfies the compatibility condition
\begin{equation}\label{gas phase compatibility condition}
\int \frac{g_g-f_g}{\tau_g} \bm{\psi} \text{d}\Xi=0,
\end{equation}
where $\bm{\psi}=\left(1,\textbf{u},\displaystyle \frac{1}{2}(\textbf{u}^2+\bm{\xi}^2)\right)^T$, the internal variables $\bm{\xi}^2=\xi_1^2+...+\xi_K^2$, and $\text{d}\Xi=\text{d}\textbf{u}\text{d}\bm{\xi}$.

For Eq.\eqref{gas phase kinetic equ without acce}, the integral solution of $f$ at the cell interface can be written as,
\begin{equation}\label{gas phase equ_integral1}
f(\textbf{x},t,\textbf{u},\bm{\xi})=\frac{1}{\tau}\int_0^t g(\textbf{x}',t',\textbf{u},\bm{\xi})e^{-(t-t')/\tau}\text{d}t'\\
+e^{-t/\tau}f_0(\textbf{x}-\textbf{u}t,\textbf{u},\bm{\xi}),
\end{equation}
where $\textbf{x}'=\textbf{x}+\textbf{u}(t'-t)$ is the trajectory of particles, $f_0$ is the initial gas distribution function at time $t=0$, and $g$ is the corresponding equilibrium state. The initial NS gas distribution function $f_0$ in Eq.\eqref{gas phase equ_integral1} can be constructed as
\begin{equation}\label{gas phase equ_f0}
f_0=f_0^l(\textbf{x},\textbf{u})H(x)+f_0^r(\textbf{x},\textbf{u})(1-H(x)),
\end{equation}
where $H(x)$ is the Heaviside function, $f_0^l$ and $f_0^r$ are the
initial gas distribution functions on the left and right side of one cell interface.
More specifically, the initial gas distribution function $f_0^k$, $k=l,r$, is constructed as
\begin{equation*}
f_0^k=g^k\left(1+\textbf{a}^k \cdot \textbf{x}-\tau(\textbf{a}^k \cdot \textbf{u}+A^k)\right),
\end{equation*}
where $g^l$ and $g^r$ are the Maxwellian distribution functions on the left and right hand sides of a cell interface, and they can be determined by the corresponding conservative variables $\textbf{W}^l$ and $\textbf{W}^r$. The coefficients $\textbf{a}^l=\left[a^l_1, a^l_2, a^l_3\right]^T$, $\textbf{a}^r=\left[a^r_1, a^r_2, a^r_3\right]^T$, are related to the spatial derivatives in normal and tangential directions, which can be obtained from the corresponding derivatives of the initial macroscopic variables,
\begin{equation*}
\left\langle a^l_i\right\rangle=\partial \textbf{W}^l/\partial x_i,
\left\langle a^r_i\right\rangle=\partial \textbf{W}^r/\partial x_i,
\end{equation*}
where $i=1,2,3$, and $\left\langle...\right\rangle$ means the moments of the Maxwellian distribution functions,
\begin{align*}
\left\langle...\right\rangle=\int \bm{\psi}\left(...\right)g\text{d}\Xi.
\end{align*}
Based on the Chapman-Enskog expansion, the non-equilibrium part of the distribution function satisfies,
\begin{equation*}
\left\langle \textbf{a}^l \cdot\textbf{u}+A^l\right\rangle = 0,~
\left\langle \textbf{a}^r \cdot\textbf{u}+A^r\right\rangle = 0,
\end{equation*}
and therefore the coefficients $A^l$ and $A^r$ can be fully determined. The equilibrium state $g$ around the cell interface is modeled as,
\begin{equation}\label{gas phase equ_g}
g=g_0\left(1+\overline{\textbf{a}}\cdot\textbf{x}+\bar{A}t\right),
\end{equation}
where $\overline{\textbf{a}}=\left[\overline{a}_1, \overline{a}_2, \overline{a}_3\right]^T$, $g_0$ is the local equilibrium of the cell interface. More specifically, $g$ can be determined by the compatibility condition,
\begin{align*}
\int\bm{\psi} g_{0}\text{d}\Xi=\textbf{W}_0
&=\int_{u>0}\bm{\psi} g^{l}\text{d}\Xi+\int_{u<0}\bm{\psi} g^{r}\text{d}\Xi, \nonumber \\
\int\bm{\psi} \overline{a_i} g_{0}\text{d}\Xi=\partial \textbf{W}_0/\partial x_i
&=\int_{u>0}\bm{\psi} a^l_i g^{l}\text{d}\Xi+\int_{u<0}\bm{\psi} a^r_i g^{r}\text{d}\Xi,
\end{align*}
$i=1,2,3$, and
\begin{equation*}
\left\langle \overline{\textbf{a}} \cdot \textbf{u}+\bar{A}\right\rangle = 0.
\end{equation*}
After determining all parameters in the initial gas distribution function $f_0$ and the equilibrium state $g$, substituting Eq.\eqref{gas phase equ_f0} and Eq.\eqref{gas phase equ_g} into Eq.\eqref{gas phase equ_integral1}, the time-dependent distribution function $f(\textbf{x}, t, \textbf{u},\bm{\xi})$ at a cell interface can be expressed as,
\begin{align}\label{gas phase equ_finalf}
f(\textbf{x}, t, \textbf{u},\bm{\xi})
&=c_1 g_0+ c_2 \overline{\textbf{a}}\cdot\textbf{u}g_0 +c_3 {\bar{A}} g_0\nonumber\\
&+\left[c_4 g^r +c_5 \textbf{a}^r\cdot\textbf{u} g^r + c_6 A^r g^r\right] (1-H(u)) \\
&+\left[c_4 g^l +c_5 \textbf{a}^l\cdot\textbf{u} g^l + c_6 A^l g^l\right] H(u) \nonumber.
\end{align}
with coefficients,
\begin{align*}
c_1 &= 1-e^{-t/\tau}, \\
c_2 &= \left(t+\tau\right)e^{-t/\tau}-\tau, \\
c_3 &= t-\tau+\tau e^{-t/\tau}, \\
c_4 &= e^{-t/\tau}, \\
c_5 &= -\left(t+\tau\right)e^{-t/\tau}, \\
c_6 &= -\tau e^{-t/\tau}.
\end{align*}
Then the integrated flux over a time step can be obtained,
\begin{align}
\textbf{F}_{ij} =\int_{0}^{\Delta t} \int\textbf{u}\cdot\textbf{n}_{ij} f_{ij}(\textbf{x},t,\textbf{u},\bm{\xi})\bm{\psi}\text{d}\Xi\text{d}t,
\end{align}
where $\textbf{n}_{ij}$ is the normal vector of the cell interface.
Then, the cell-averaged conservative variables of cell $i$ can be updated as follows,
\begin{gather}
\textbf{W}_i^{n+1} = \textbf{W}_i^n
- \frac{1}{\Omega_i} \sum_{S_{ij}\in \partial \Omega_i}\textbf{F}_{ij}S_{ij},
\end{gather}
where $\Omega_i$ is the volume of cell $i$, $\partial\Omega_i$ denotes the set of interface of cell $i$, $S_{ij}$ is the area of $j$-th interface of cell $i$, $\textbf{F}_{ij}$ denotes the projected macroscopic fluxes in the normal direction, and $\textbf{W}_{g}=\left[\rho_g,\rho_g \textbf{U}_g, \rho_g E_g\right]^T$ are the cell-averaged conservative flow variables for gas phase.

The second part, $\mathcal{L}_{g2}$, is from the inter-phase interaction. The increased macroscopic variables for gas phase in 3D can be calculated as
\begin{gather}
\textbf{W}^{n+1}_g = \textbf{W}_g + \textbf{Q}\Delta t,
\end{gather}
where
\begin{gather*}
\textbf{W}_g=\left[\begin{array}{c}
\rho_g\\
\rho_g \textbf{U}_g\\
\rho_g E_g
\end{array}
\right], ~~
\textbf{Q}=\left[\begin{array}{c}
C_{\epsilon_g}\rho_g \\
C_{\epsilon_g} \rho_g \textbf{U}_g
-\frac{\epsilon_{s}\rho_{s}\left(\textbf{U}_g - \textbf{U}_s\right)}{\epsilon_g \tau_{st}}
+ \rho_g\textbf{G}\\
C_{\epsilon_g} \left(\rho_g E_g + p_g\right)
-\frac{\epsilon_{s}\rho_{s}\textbf{U}_s \cdot \left(\textbf{U}_g - \textbf{U}_s\right)}{\epsilon_g \tau_{st}}
+ \frac{3p_s}{\epsilon_g \tau_{st}}
+\rho_g \textbf{U}_g \cdot \textbf{G}
\end{array}\right],
\end{gather*}
with $C_{\epsilon_g} = -\frac{1}{\epsilon_{g}}\frac{\text{d}\epsilon_{g}}{\text{d}t}$ and $\frac{\text{d}\epsilon_{g}}{\text{d}t}=\frac{\partial \epsilon_{g}}{\partial t}+\textbf{U}_g \cdot \nabla\epsilon_{g}$. In this paper, $\frac{\partial \epsilon_{g}}{\partial t}$ is evaluated,
\begin{equation}
\frac{\partial \epsilon_{g}}{\partial t} = \frac{\epsilon_{g}^{n+1} - \epsilon_{g}^n}{\Delta t}.
\end{equation}
Here $\nabla\epsilon_{g}$ is the cell-averaged volume fraction gradient of gas phase in the cell. For example, $\frac{\partial \epsilon_{g}}{\partial x}$ is calculated by,
\begin{equation}
\frac{\partial \epsilon_{g,i}}{\partial x} = \frac{\epsilon_{g,i+\frac{1}{2}} - \epsilon_{g,i-\frac{1}{2}}}{\Delta x},
\end{equation}
where $\epsilon_{g,i-\frac{1}{2}}$ and $\epsilon_{g,i+\frac{1}{2}}$ are volume fractions of gas phase at the left and right interface of cell $i$, which can be obtained from the reconstructed $\epsilon_{s}$ according to $\epsilon_{s} + \epsilon_{g} = 1$.
Now the update for the gas phase in one time step has been finished.

\begin{figure}[htbp]
	\centering
	\subfigure{
		\includegraphics[height=10.5cm]{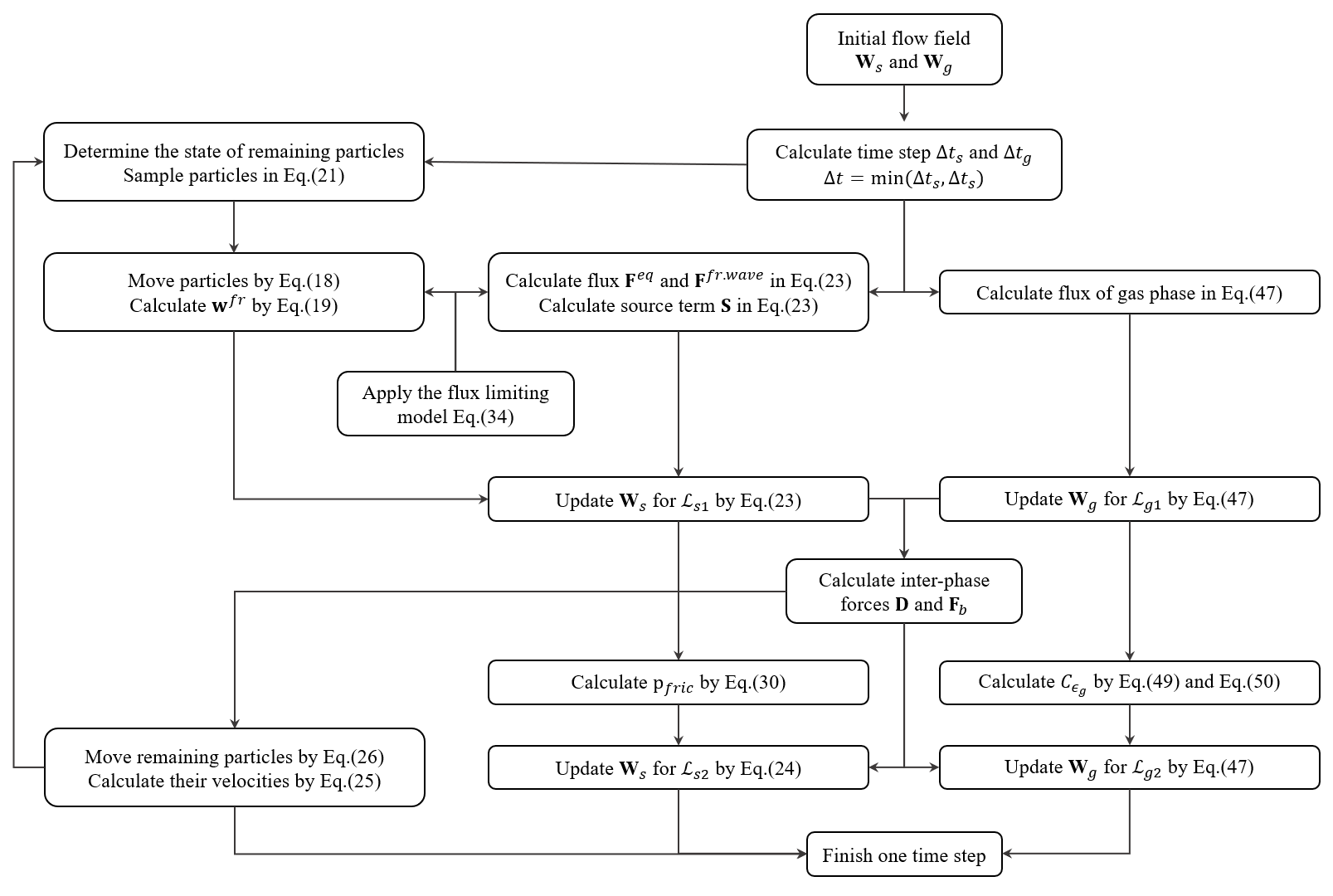}}
	\caption{The flow chart of GKS-UGKWP method.}
	\label{Flow chart}
\end{figure}

Finally, the algorithm of GKS-UGKWP method for the gas-particle two phase flow is summarized in Figure \ref{Flow chart}.

\section{Numerical experiments}
\subsection{Interaction of a shock wave with dense particle layer}
\begin{figure}[htbp]
	\centering
	\subfigure{
		\includegraphics[height=3.5cm]{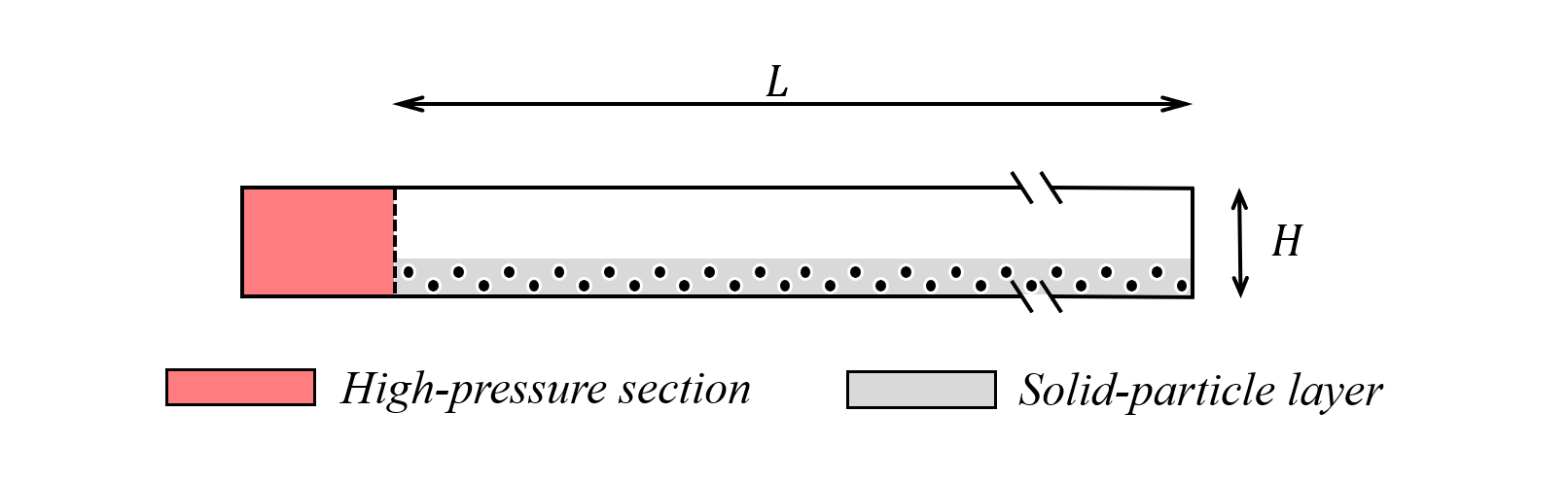}}
	\caption{Sketch of initial condition.}
	\label{Sketch of the interaction of shock and particle layer problem}
\end{figure}

The interaction of a shock wave with a dense particle layer will generate complicated particles' behavior \cite{Gasparticle-shock-particle-layer-kosinski2005dust, Gasparticle-shock-particle-layer-step-shimura2018two}, which brings challenges to a numerical scheme. The problem in \cite{Gasparticle-shock-particle-layer-kosinski2005dust} is tested by GKS-UGKWP in this section. Figure \ref{Sketch of the interaction of shock and particle layer problem} presents the initial configuration of the test case. The computational domain is a channel with size $L\times H=0.1m\times0.005m$, which is covered by $250\times20$ uniform rectangular mesh. The initial height of the dense particle layer in the channel is $0.001m$, and the volume fraction is 0.5. The layer is composed of solid particles with density $1000kg$ and diameter $90\mu m$. Initially, the gas in the channel is standard atmospheric condition. Next to the particle layer, there is a high pressure gas region with $4$bar, which will generate a shock wave after the diaphragm is removed at the beginning of calculation.

\begin{figure}[htbp]
	\centering
	\subfigure{
		\includegraphics[height=1.0cm]{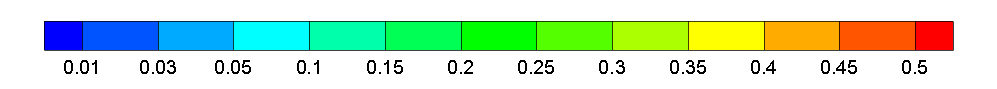}}
	\quad
	\subfigure{
		\includegraphics[height=2.0cm]{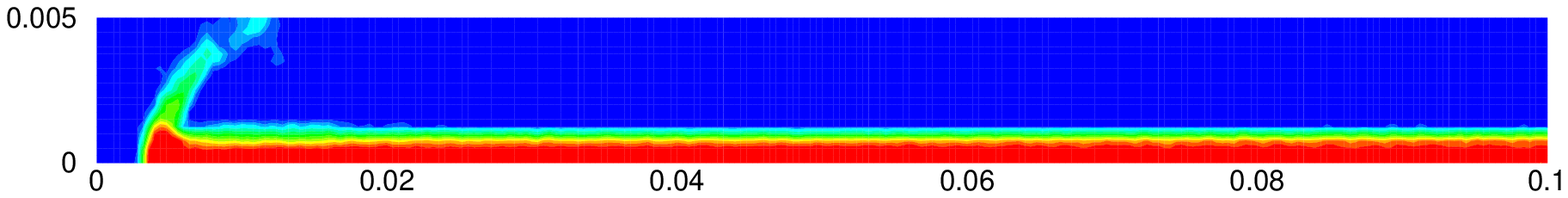}}
	\quad
	\subfigure{
		\includegraphics[height=2.0cm]{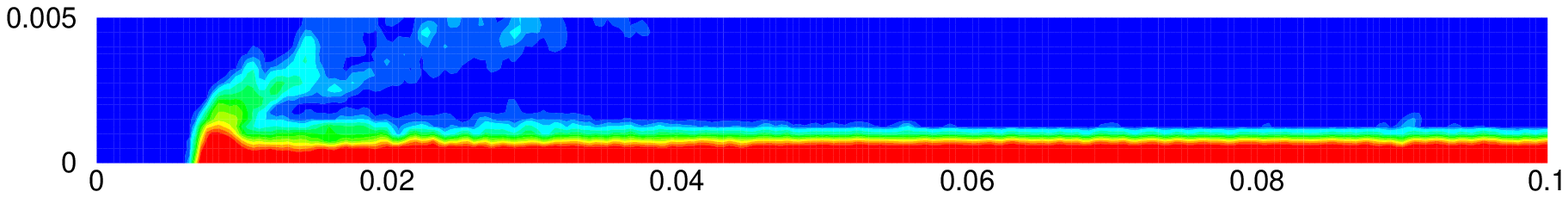}}
	\quad
	\subfigure{
		\includegraphics[height=2.0cm]{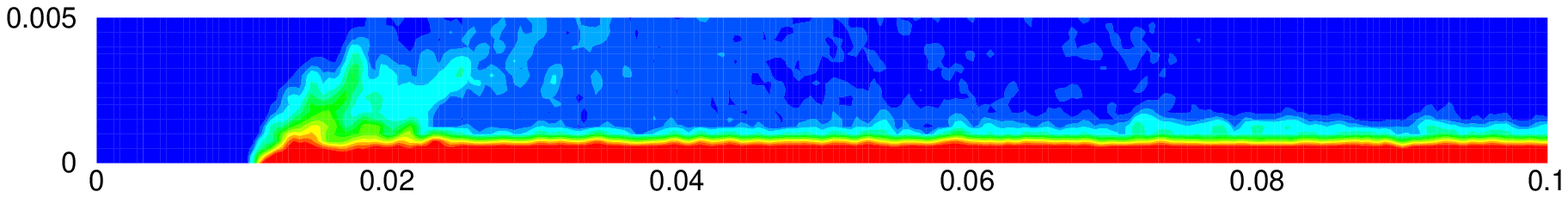}}
	\quad
	\subfigure{
		\includegraphics[height=2.0cm]{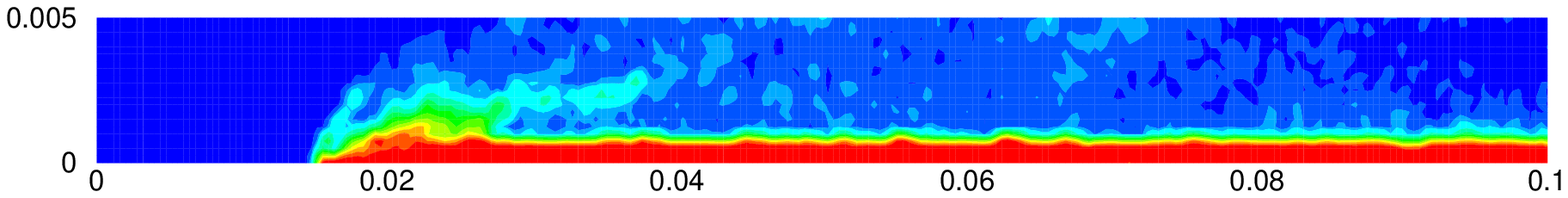}}	
	\caption{Particle phase volume fraction at $t=0.3ms$, $t=0.6ms$, $t=1.0ms$ and $t=1.4ms$.}
	\label{Shock particle layer eps}
\end{figure}

\begin{figure}[htbp]
	\centering
	\subfigure{
		\includegraphics[height=5.5cm]{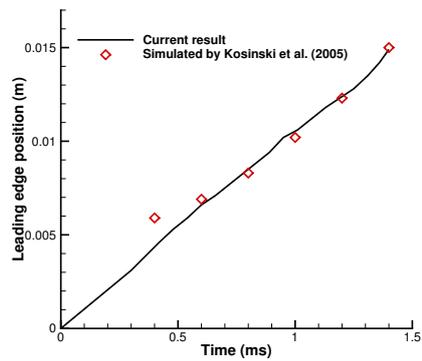}}
	\caption{The leading edge position of dense particle layer at different time.}
	\label{Shock particle layer leading edge}
\end{figure}

The post-shock snapshots of solid-particle phase volume fraction are shown in Figure \ref{Shock particle layer eps}. After the shock wave passes, more and more particles in the dense layer will be lifted upward and therefore a ``particle stream" is formed at the leading section of the layer. The lifted particles will be accelerated by the gas flow and move backward, and then solid particles are dispersed in the channel. These particle behaviors have also been observed in the previous studies \cite{Gasparticle-shock-particle-layer-kosinski2005dust, Gasparticle-shock-particle-layer-step-shimura2018two}. Since more and more particles are lifted upward and dispersed in the channel, the leading edge of the dense particle layer gradually moves backward. The changing of leading-edge position with time is shown in Figure \ref{Shock particle layer leading edge}, which agrees well with the previous study by Eulerian-Lagrangian approach \cite{Gasparticle-shock-particle-layer-kosinski2005dust}.

\begin{figure}[htbp]
	\centering
	\subfigure{
		\includegraphics[height=1.0cm]{figure/Shock-particle-layer/legend-epss.png}}
	\quad
	\subfigure{
		\includegraphics[height=2.0cm]{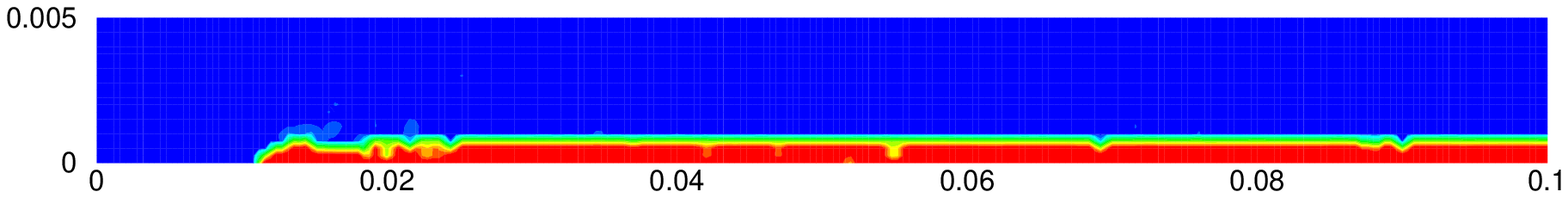}}
	\quad
	\subfigure{
		\includegraphics[height=2.0cm]{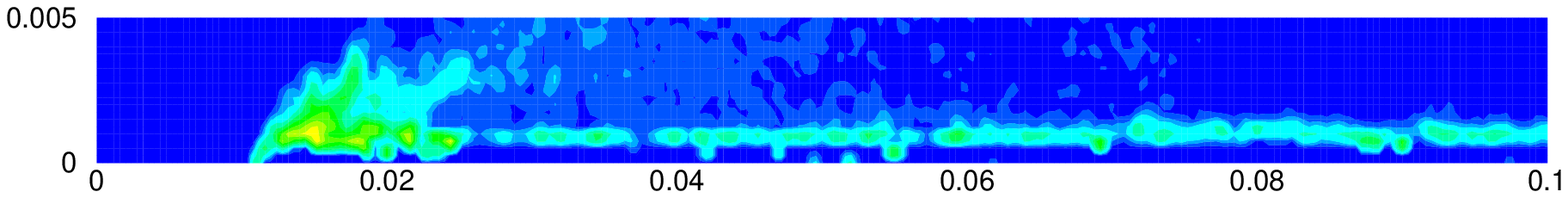}}
	\caption{UGKWP computation of solid-particle phase by wave (up) and particle (down) decompositions at $t=1.0ms$.}
	\label{Shock particle layer wave and particle in 1.0ms}
\end{figure}

Figure \ref{Shock particle layer wave and particle in 1.0ms} shows the wave and particle decompositions from UGKWP at $t=1.0ms$. For the dense particle layer region, e.g., the zone near bottom wall, inter-particle collisions play the key role in the evolution due to the high solid concentration. In UGKWP, no particle will be sampled there and only wave is used for the evolution of particle flow, such as the automatic fluid approach. However, for the dilute particle region in the up part of the channel, the non-equilibrium particle transport appears and particles are sampled and tracked in UGKWP. Therefore, the UGKWP can adapt to different flow physics consistently.
In addition, the percentage of sampled particles in UGKWP is fully determined by local flow condition, which is not artificially pre-defined.
The above results indicate that UGKWP unifies the approaches for the equilibrium and non-equilibrium transport seamlessly, and provides an efficient method for the multiscale flow simulation.

\subsection{Horizontal pneumatic conveying}
Pneumatic conveying is a widely-used technique for the transportation of bulk solid particles by gas flow in the pipe or channel. The advantage of pneumatic conveying includes design flexibility, working safety, and low maintenance cost, etc \cite{Gasparticle-book-fan1999principles}. Under different conditions, the solid phase will show different flow patterns. Here, a horizontal pneumatic conveying problem will be tested by GKS-UGKWP to check its ability to recover the typical flow patterns. The flow conditions, including inlet gas velocity $U_{g,in}$, inlet solid mass flow rate $G_{s,in}$ and gas pressure gradient $\Delta p/L$, obtained from the experiment \cite{Gasparticle-pneumatic-converying-rao2001electrical}, are employed in the simulation. The solid particles used in the experiment have the following physical properties: density $1683kg/m^3$ and diameter $3.01mm$. The computational domain is a two-dimensional horizontal channel with size $4m\times0.04m$, covered by $800\times8$ uniform rectangular mesh. Three typical cases, disperse flow pattern, settle flow pattern, and slug flow pattern, are tested, and the corresponding experimental measurement data are listed in Table \ref{pneumatic conveying problem three cases table}. Initially, the gas with inlet velocity $U_{g,in}$ flows into the channel from the left boundary; the solid particles are carried by gas flow and uniformly fed into the channel with mass flow rate $G_{s,in}$ through the left boundary; at the right boundary solid particles are free to leave. The atmospheric pressure boundary is employed for gas phase at the right boundary, while higher gas pressure is imposed at the left boundary according to the pressure gradient $\Delta p/L$ given in Table \ref{pneumatic conveying problem three cases table}.

\begin{table}[h]
	\caption{Simulation conditions from experimental measurement \cite{Gasparticle-pneumatic-converying-rao2001electrical}.}
	\vspace{2pt}
	\small
	\centering
	\setlength{\tabcolsep}{3.4mm}{
		\begin{tabular}{ccccc}\toprule[1pt]		
			& $U_{g,in}$ $\left(m/s\right)$  & $G_{s,in}$ $\left(kg/m^2\cdot s\right)$  & $\Delta p/L$ $\left(Pa/m\right)$ & \textit{Flow pattern} \\ \hline
			Case 1 & 28.6 & 71.4 & 271.4 & disperse flow\\
			Case 2 & 15.6 & 17.2 & 454.0 & settle flow\\
			Case 3 & 10.4 & 21.1 & 855.6 & slug flow \\
			\bottomrule[1pt]		
	\end{tabular}}
	\label{pneumatic conveying problem three cases table}
\end{table}

\begin{figure}[htbp]
	\centering
	\subfigure{
		\includegraphics[height=0.7cm]{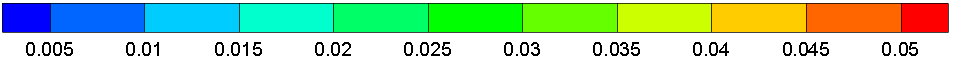}}
	\quad	
	\subfigure{
		\includegraphics[height=1.8cm]{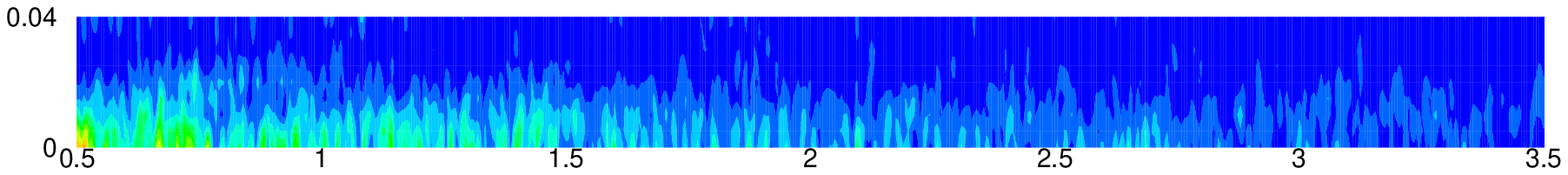}}	
	\caption{The snapshot of solid phase volume fraction $\epsilon_{s}$ of Case 1, disperse flow pattern, at $t=6.0s$.}
	\label{pneumatic conveying problem disperse flow}
\end{figure}

\begin{figure}[htbp]
	\centering
	\subfigure{
		\includegraphics[height=0.7cm]{figure/pneumatic-converying/disperse-flow-t6d0-legend}}
	\quad	
	\subfigure{
		\includegraphics[height=1.8cm]{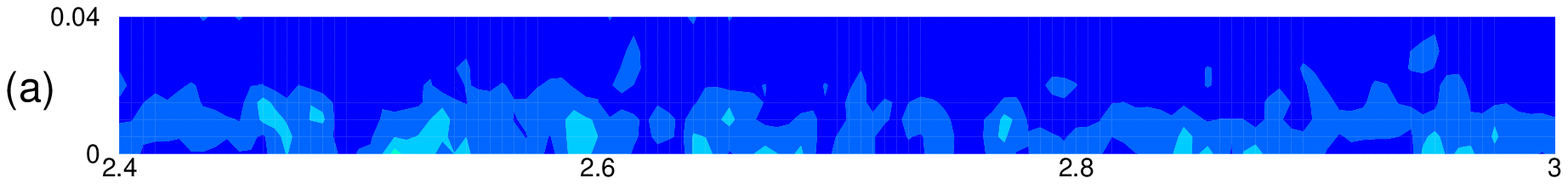}}
	\quad
	\subfigure{
		\includegraphics[height=1.8cm]{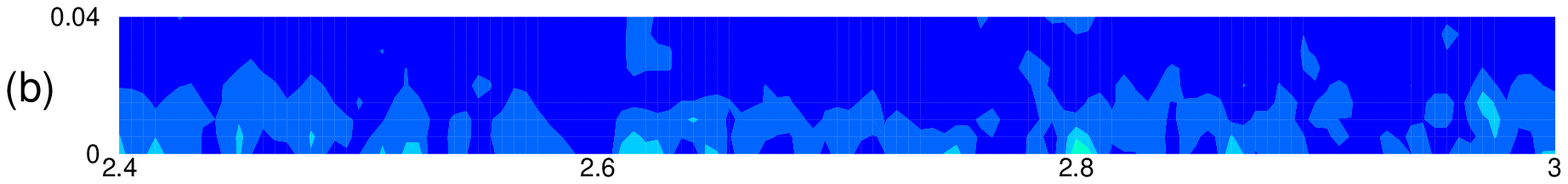}}	
	\quad
	\subfigure{
		\includegraphics[height=1.8cm]{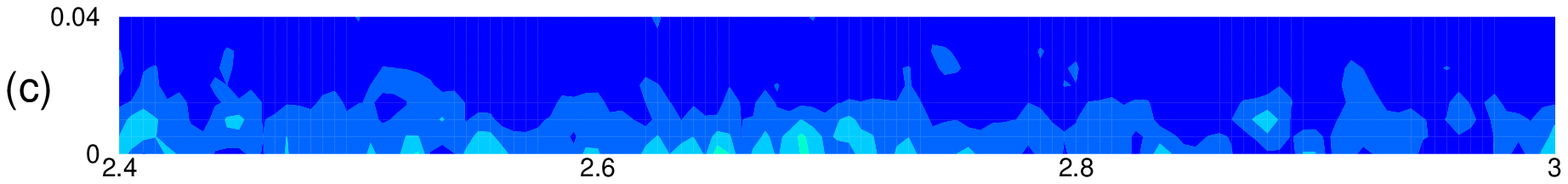}}	
	\caption{The enlarged snapshots of solid phase volume fraction $\epsilon_{s}$ in the local region $2.4m\sim3.0m$ of Case 1 at different times: (a) $t=5.0s$, (b)$t=5.5s$, (c)$t=6.0s$.}
	\label{pneumatic conveying problem disperse flow enlarge}
\end{figure}

\begin{figure}[htbp]
	\centering
	\subfigure{
		\includegraphics[height=0.7cm]{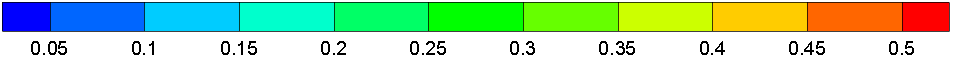}}
	\quad	
	\subfigure{
		\includegraphics[height=1.8cm]{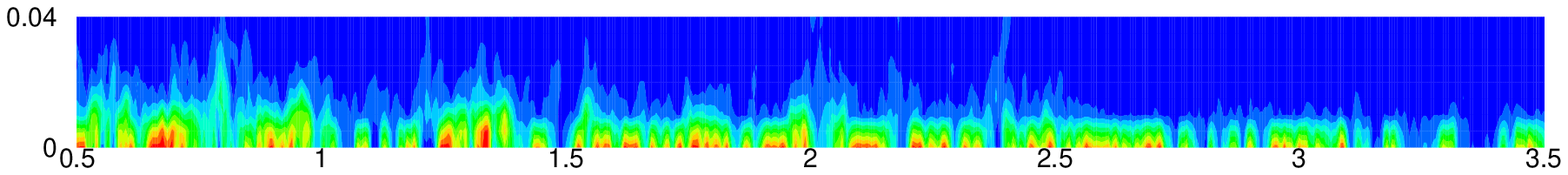}}	
	\caption{The snapshot of solid phase volume fraction $\epsilon_{s}$ of Case 2, settle flow pattern, at $t=6.0s$. The whole computation domain $4m\times0.04m$ is shown.}
	\label{pneumatic conveying problem settle flow}
\end{figure}

\begin{figure}[htbp]
	\centering
	\subfigure{
		\includegraphics[height=0.7cm]{figure/pneumatic-converying/settle-flow-t6d0-legend}}
	\quad	
	\subfigure{
		\includegraphics[height=1.8cm]{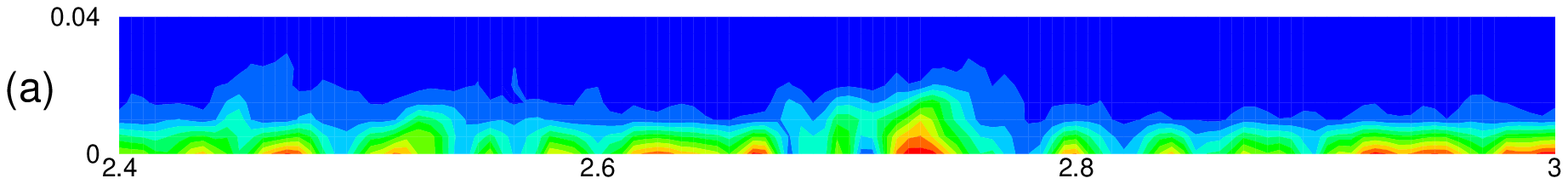}}
	\quad
	\subfigure{
		\includegraphics[height=1.8cm]{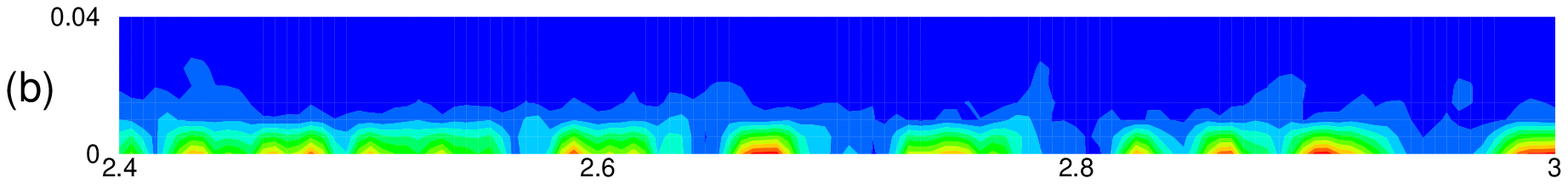}}	
	\quad
	\subfigure{
		\includegraphics[height=1.8cm]{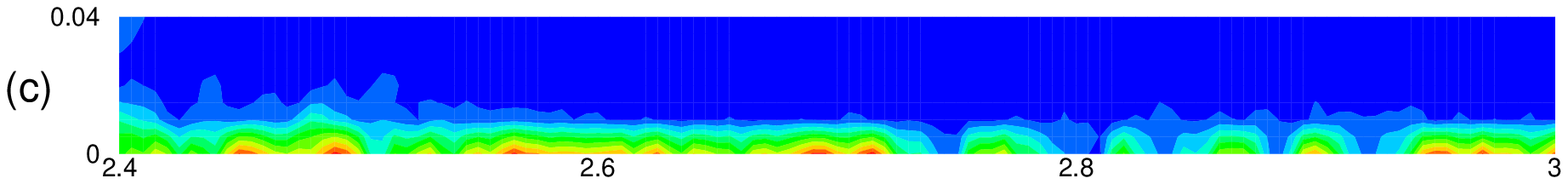}}	
	\caption{The enlarged snapshots of solid phase volume fraction $\epsilon_{s}$ in the local region $2.4m\sim3.0m$ of Case 2 at different times: (a) $t=5.0s$, (b)$t=5.5s$, (c)$t=6.0s$.}
	\label{pneumatic conveying problem settle flow enlarge}
\end{figure}

\begin{figure}[htbp]
	\centering	
	\subfigure{
		\includegraphics[height=0.7cm]{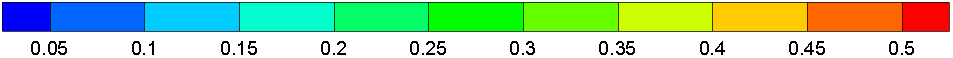}}
	\quad
	\subfigure{
		\includegraphics[height=1.8cm]{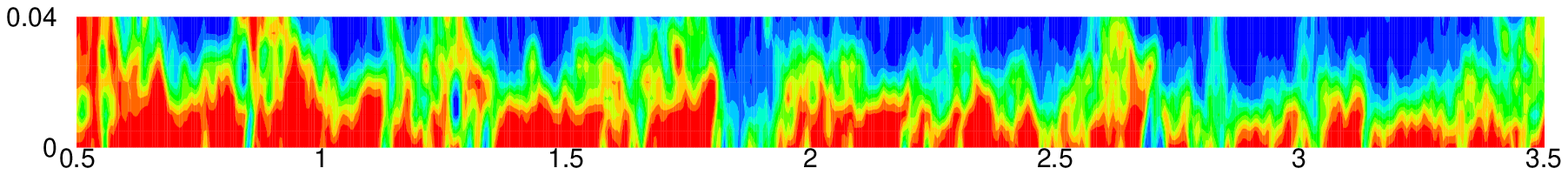}}	
	\caption{The snapshot of solid phase volume fraction $\epsilon_{s}$ of Case 3, slug flow pattern, at $t=6.0s$. The whole computation domain $4m\times0.04m$ is shown.}
	\label{pneumatic conveying problem slug flow}
\end{figure}

\begin{figure}[htbp]
	\centering
	\subfigure{
		\includegraphics[height=0.7cm]{figure/pneumatic-converying/slug-flow-t6d0-legend}}
	\quad	
	\subfigure{
		\includegraphics[height=1.8cm]{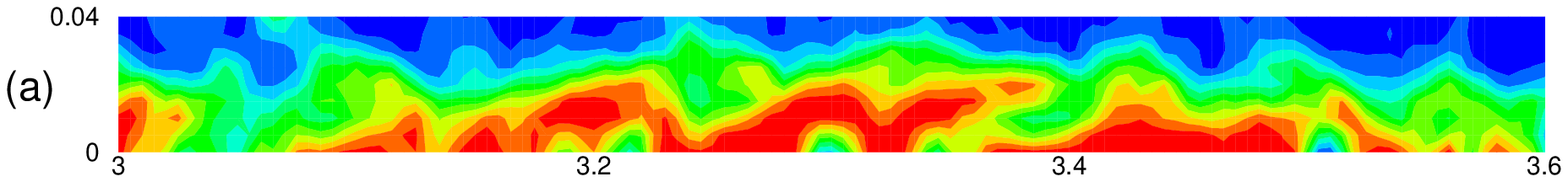}}
	\quad
	\subfigure{
		\includegraphics[height=1.8cm]{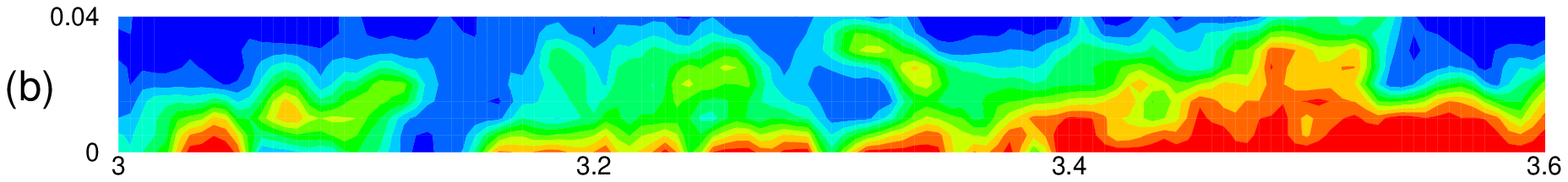}}	
	\quad
	\subfigure{
		\includegraphics[height=1.8cm]{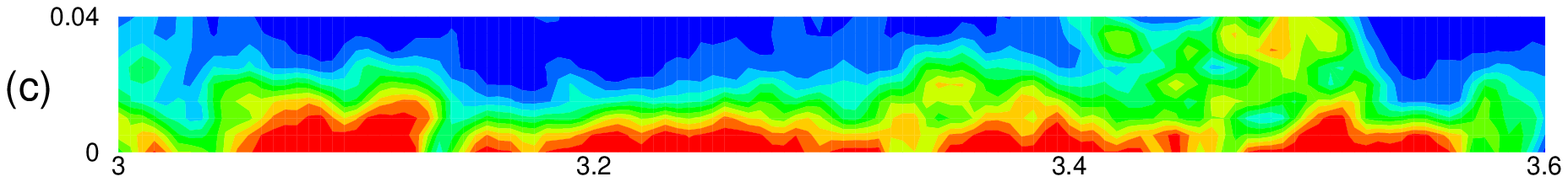}}	
	\caption{The enlarged snapshots of solid phase volume fraction $\epsilon_{s}$ in the local region $3.0m\sim3.6m$ of Case 3 at different times: (a) $t=5.0s$, (b)$t=5.5s$, (c)$t=6.0s$.}
	\label{pneumatic conveying problem slug flow enlarge}
\end{figure}

For Case 1, the snapshot of solid phase volume fraction $\epsilon_{s}$ in the region $0.5m\sim3.5m$ at $t=6.0s$ is shown in Figure \ref{pneumatic conveying problem disperse flow}, and the enlarged snapshots at different times are presented in Figure \ref{pneumatic conveying problem disperse flow enlarge}. The typical disperse flow pattern is observed: solid phase is in dilute flow region (the solid volume fraction $\epsilon_{s}$ of most areas in the channel is lower than 0.01); solid particles move downstream carried by gas flow and solid particle concentration is relatively higher at the channel bottom than the up zone due to the effect of gravity.
For Case 2, the snapshot of solid volume fraction $\epsilon_{s}$ in the channel at $t=6.0s$ and the enlarged snapshots in the local region $2.4m\sim3.0m$ are shown in Figure \ref{pneumatic conveying problem settle flow} and Figure \ref{pneumatic conveying problem settle flow enlarge}, respectively. In Case 2, a settled layer of solid particles with $\epsilon_{s}$ around 0.3 are formed along the channel bottom; while dilute solids flow is observed above this settle layer. It is the typical structure for settle flow pattern, or called stratified flow pattern.
Finally, the snapshot of solid volume fraction $\epsilon_{s}$ in the channel at $6.0s$ for Case 3 and the local enlarged snapshots at $5.0s, 5.5s,$ and $6.0s$ are presented in Figure \ref{pneumatic conveying problem slug flow} and Figure \ref{pneumatic conveying problem slug flow enlarge}. Compared with the flow conditions of Case 2, Case 3 has a lower inlet gas velocity and a greater inlet solid particle flux, and therefore the solid concentration is generally higher in the channel. Particularly, in some zones the solid phase is in dense flow on the whole cross-section of the channel, which is the typical phenomenon for slug flow pattern.
In summary, the flow structures and features predicted by GKS-UGKWP are consistent with the experimental observations for three typical flow patterns, validating the feasibility and reliability of GKS-UGKWP for this kind of problems.

\subsection{Bubble formation in fluidized bed}
\begin{figure}[htbp]
	\centering
	\subfigure{
		\includegraphics[height=5.5cm]{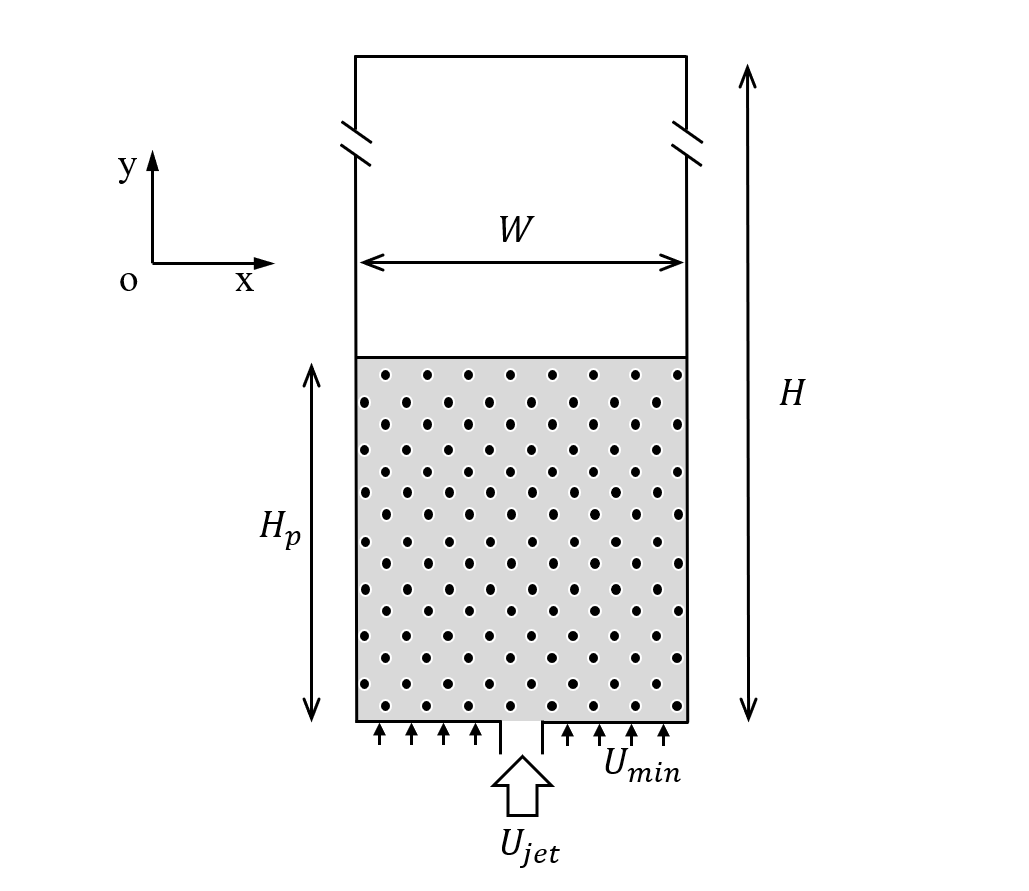}}
	\caption{Sketch of bubble formation in fluidized bed.}
	\label{Sketch of bubble formation}
\end{figure}

The fluidized bed is widely used in chemical industry to enhance chemical reactions, solids separation, heat transfer, etc.
In this problem, the initial stage of bubble formation in a fluidized bed is simulated, and the detailed description of this experiment can refer to \cite{Gasparticle-fluidized-single-bubble-nieuwland1996bubble}.
The sketch of this problem is shown in Figure \ref{Sketch of bubble formation}. The computational domain $W\times H$ is $0.57m\times1.0m$, and $76\times120$ uniform rectangular mesh is used. An orifice with width $0.02m$ exists at the bottom center. The height of particle bed $H_p$ is $0.5m$, and above this particle bed is free board used for the expansion of particle bed. The bed consists of solid particles with density $3060kg/m^3$ and diameter $285\mu m$. The initial solid volume fraction $\epsilon_s$ is set as 0.5, which is smaller than $\epsilon_{s,max}$ taken as 0.6 in this case. This is based on the condition that the initial particle bed has reached a minimum fluidization state before blowing upward gas flow into the particle bed. Initially, the jet with $U_{jet}=10.0m/s$ blows into the particle bed through the orifice, while the gas with the minimum fluidization velocity $U_{min}=0.08m/s$ flows into the particle bed at other bottom boundary region outside the center orifice. For gas phase, the up boundary is set as pressure outlet, and for the bottom boundary a higher pressure is employed with $\Delta p = 7500Pa$, which is approximated to balance the gravity by $\Delta p = \epsilon_{s}\left(\rho_s-\rho_g\right)GH_p$ as given in \cite{Gasparticle-fluidized-single-bubble-nieuwland1996bubble}. For the left and right walls, the non-slip and slip boundary condition is employed for gas phase and solid particle phase, respectively.

\begin{figure}[htbp]
	\centering
	\subfigure{
		\includegraphics[height=1.1cm]{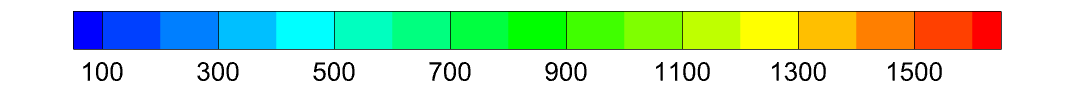}}
	\quad
	\subfigure{
		\includegraphics[height=5.0cm]{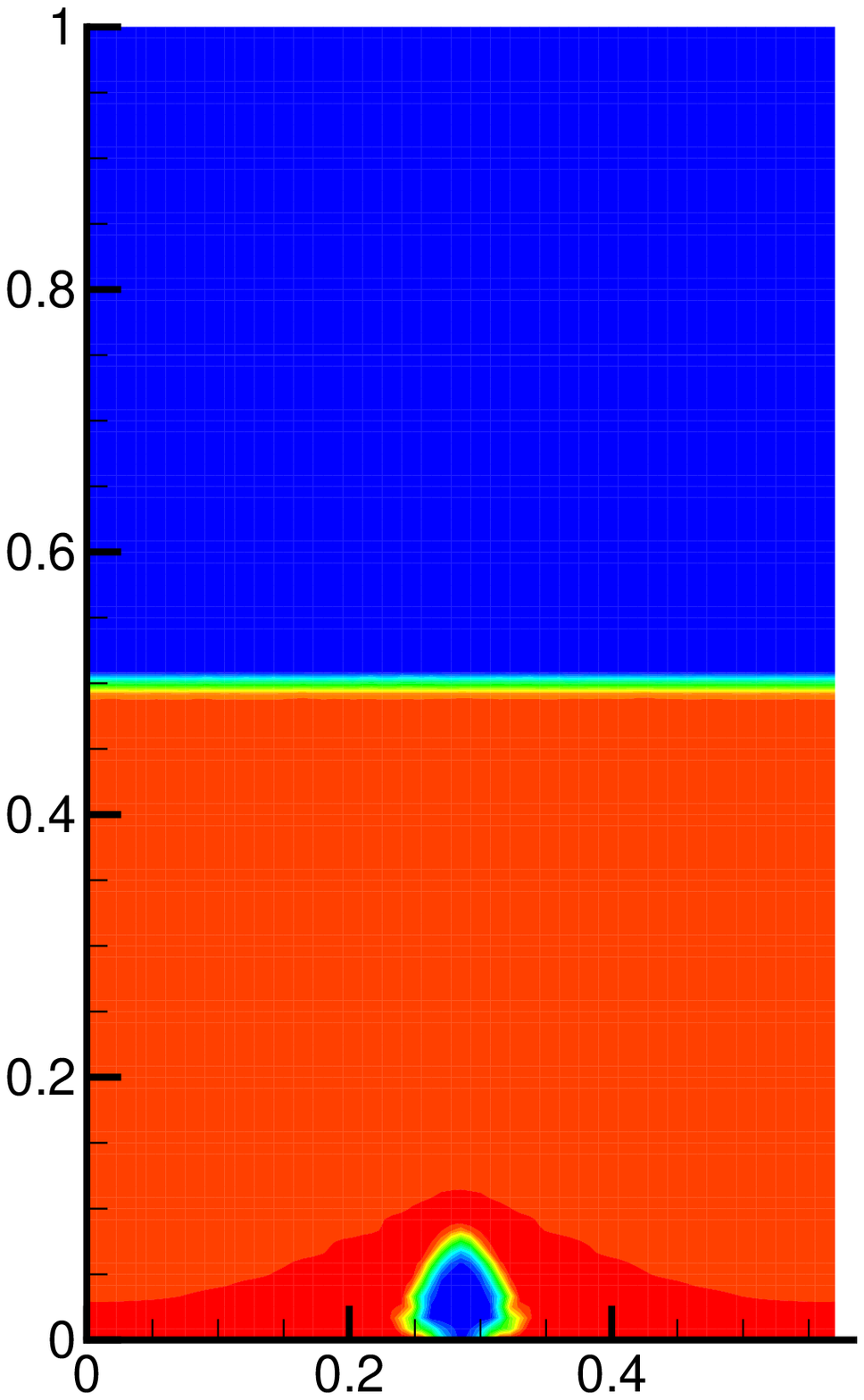}}
	\quad	
	\subfigure{
		\includegraphics[height=5.0cm]{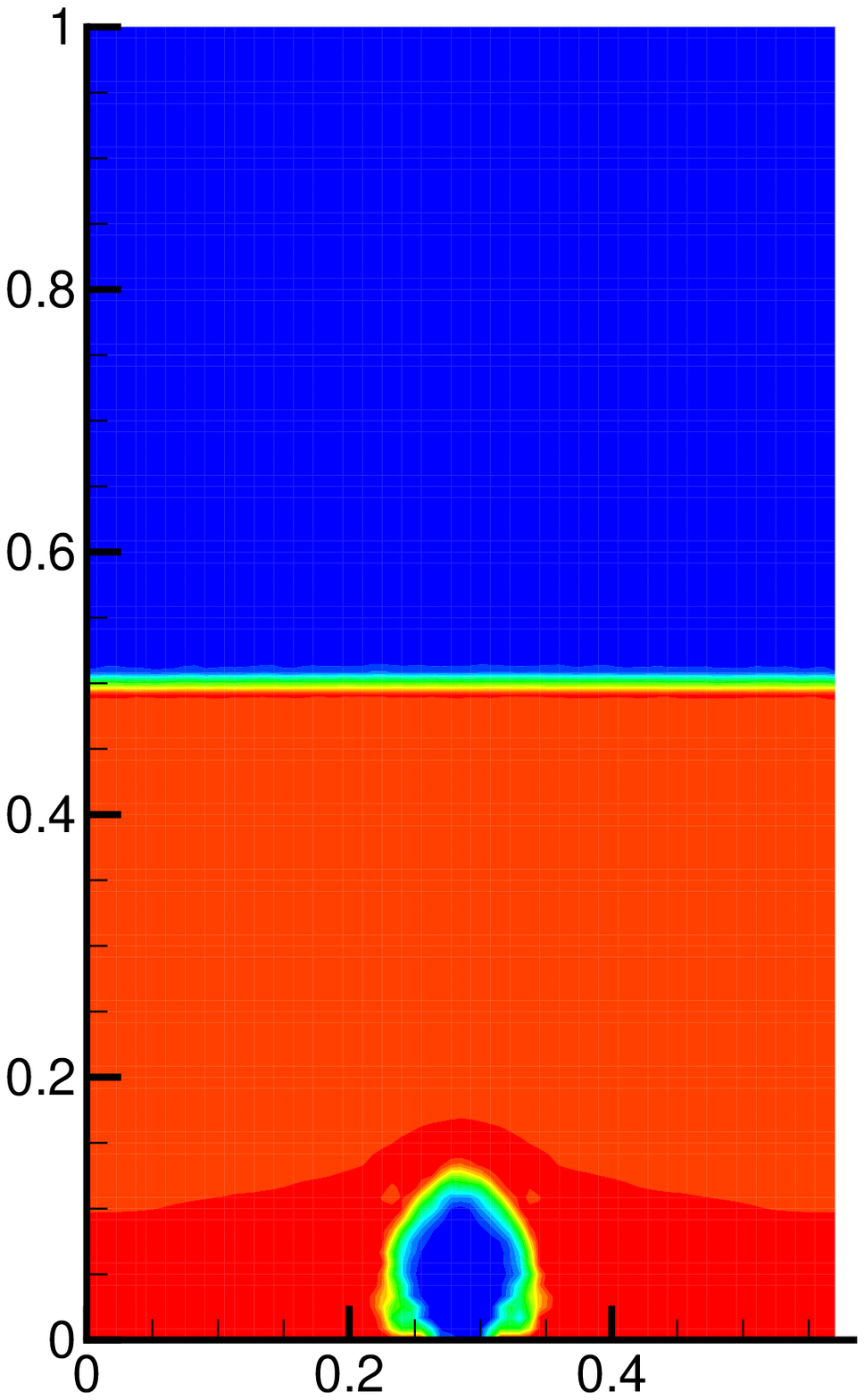}}
	\quad
	\subfigure{
		\includegraphics[height=5.0cm]{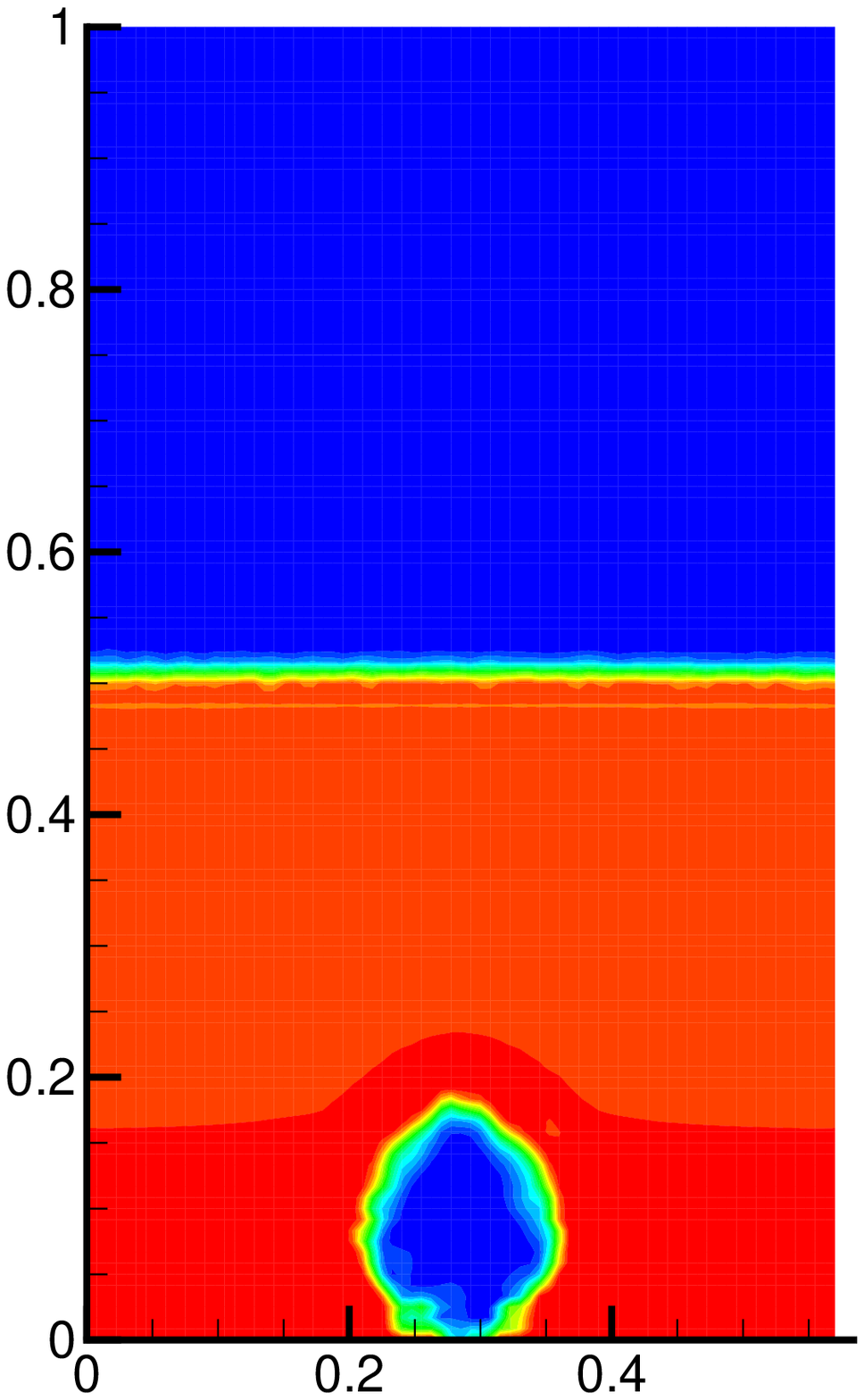}}
	\quad	
	\subfigure{
		\includegraphics[height=5.0cm]{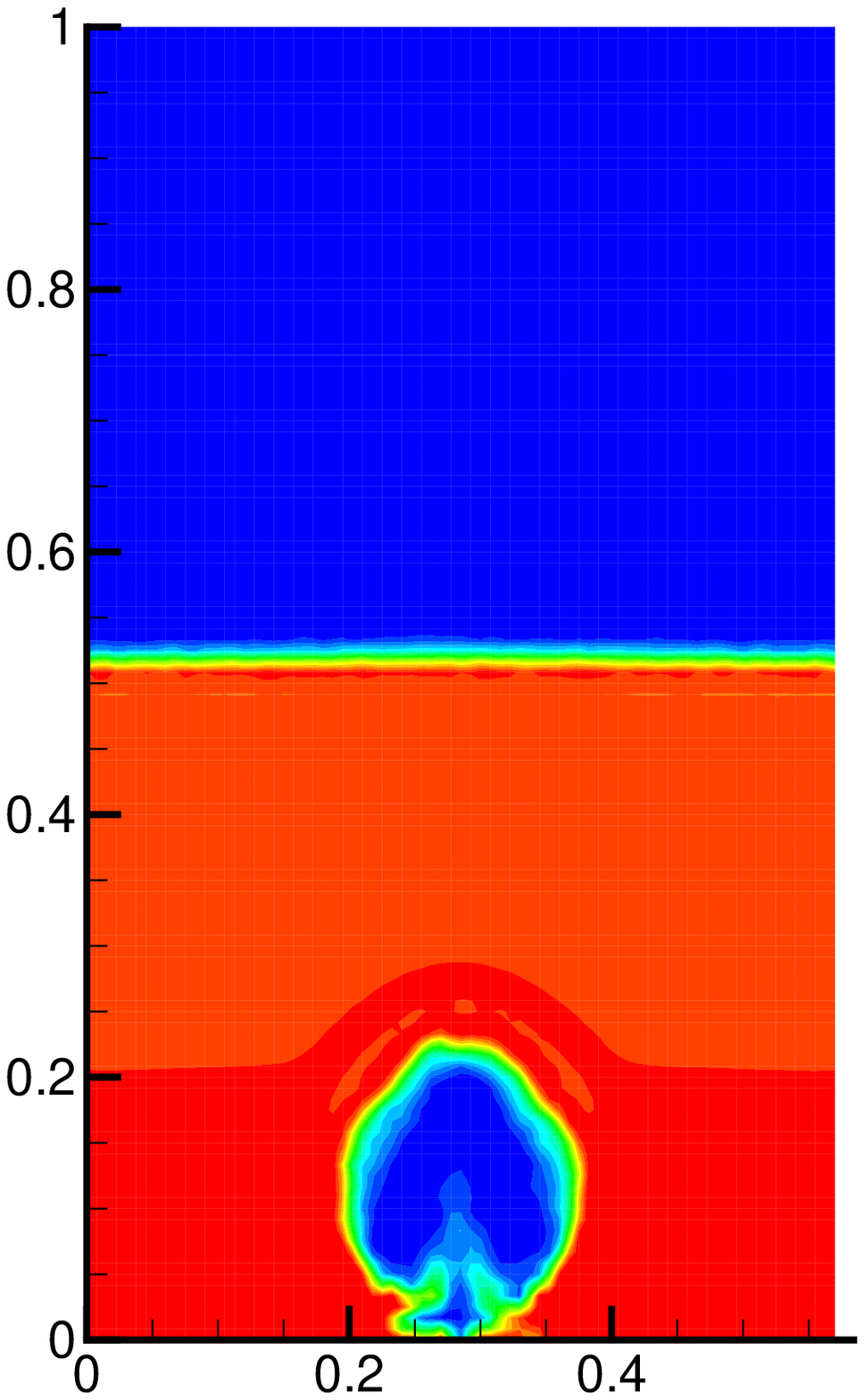}}
	\caption{Apparent density of solid particle phase during bubble formation process: from left to right are the snapshots at time $0.05s$, $0.10s$, $0.15s$, and $0.18s$.}
	\label{bubble formation apparent density}
\end{figure}

\begin{figure}[htbp]
	\centering
	\subfigure{
		\includegraphics[height=5.5cm]{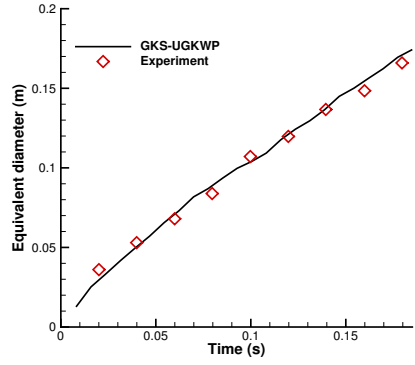}}
	\caption{Comparison of equivalent diameter $D_{e}$ obtained by GKS-UGKWP with experiment measurement.}
	\label{bubble formation ed compare}
\end{figure}

\begin{figure}[htbp]
	\centering
	\subfigure{
		\includegraphics[height=1.1cm]{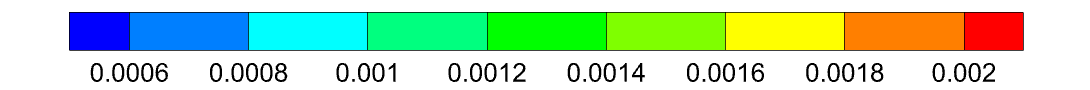}}
	\quad
	\subfigure{
		\includegraphics[height=4.0cm]{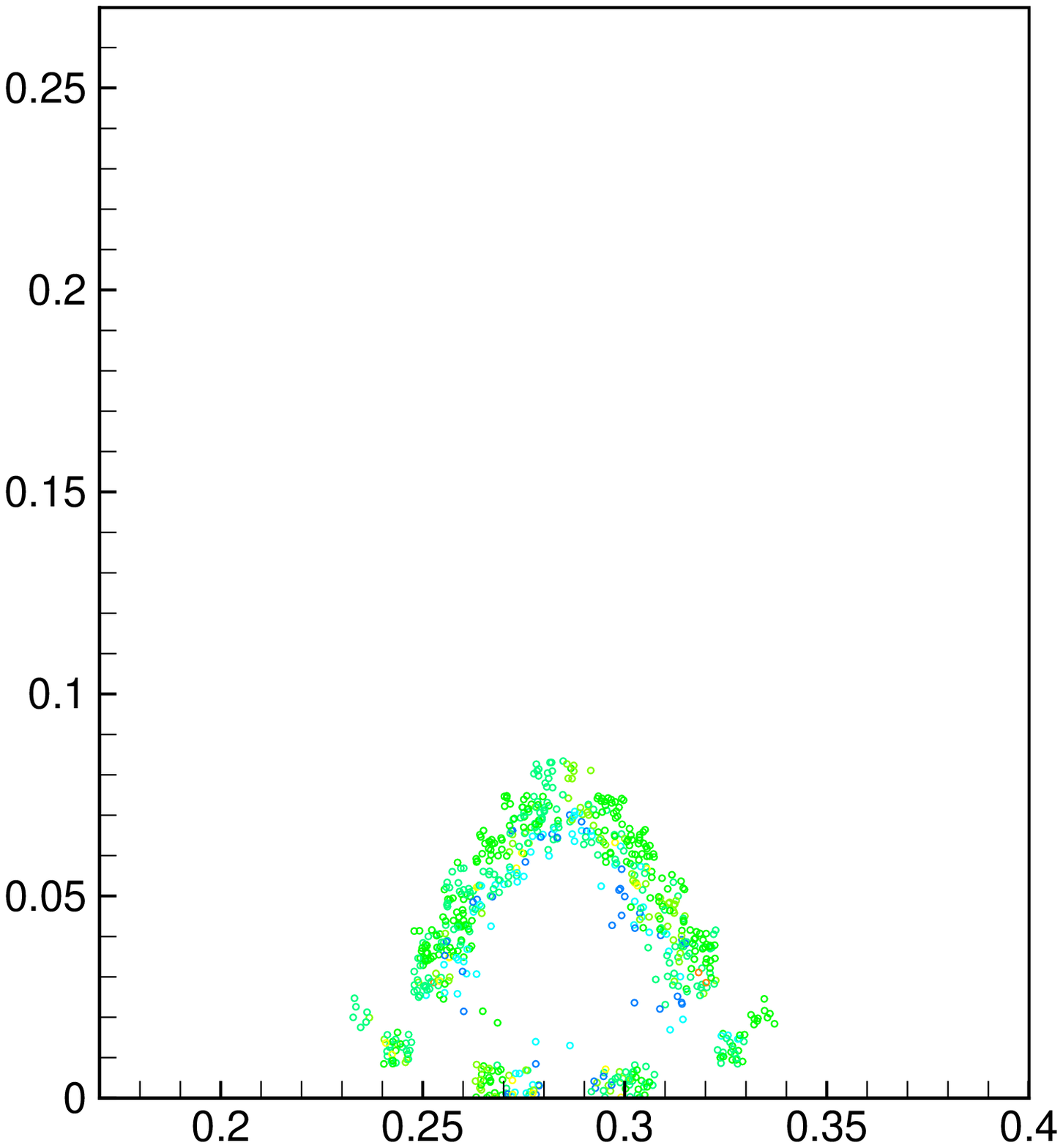}}
	\quad	
	\subfigure{
		\includegraphics[height=4.0cm]{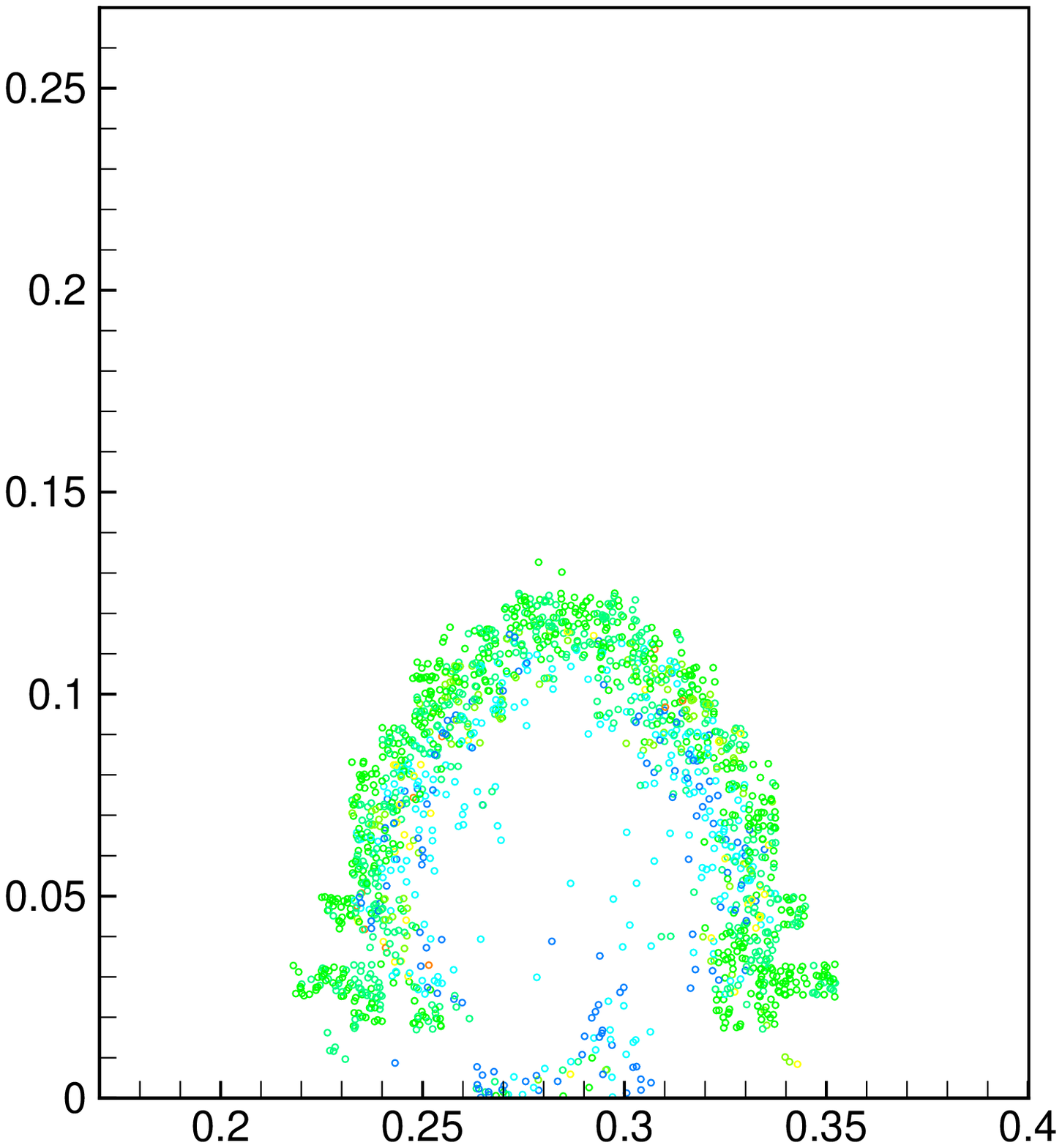}}
	\quad
	\subfigure{
		\includegraphics[height=4.0cm]{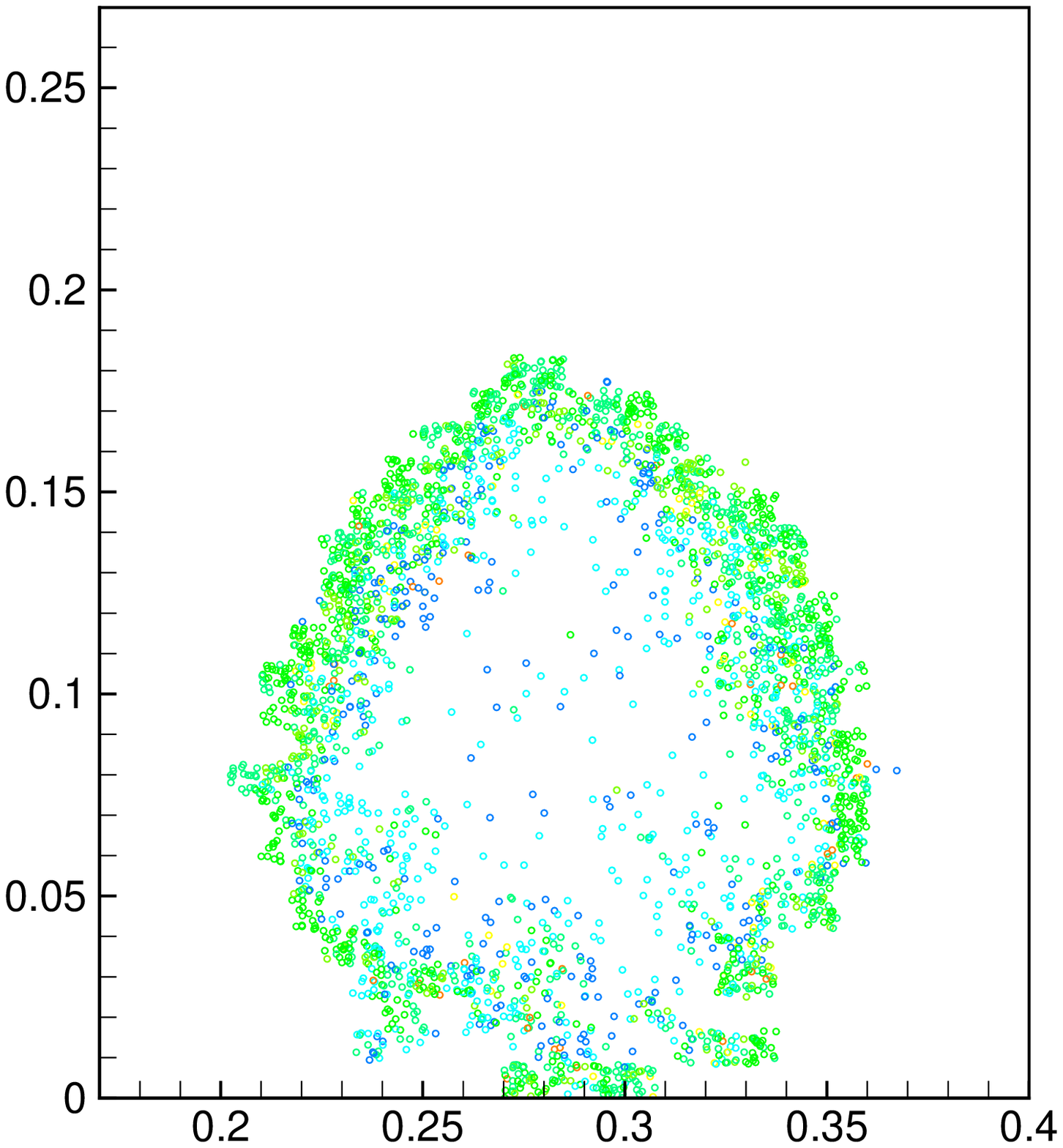}}
	\quad	
	\subfigure{
		\includegraphics[height=4.0cm]{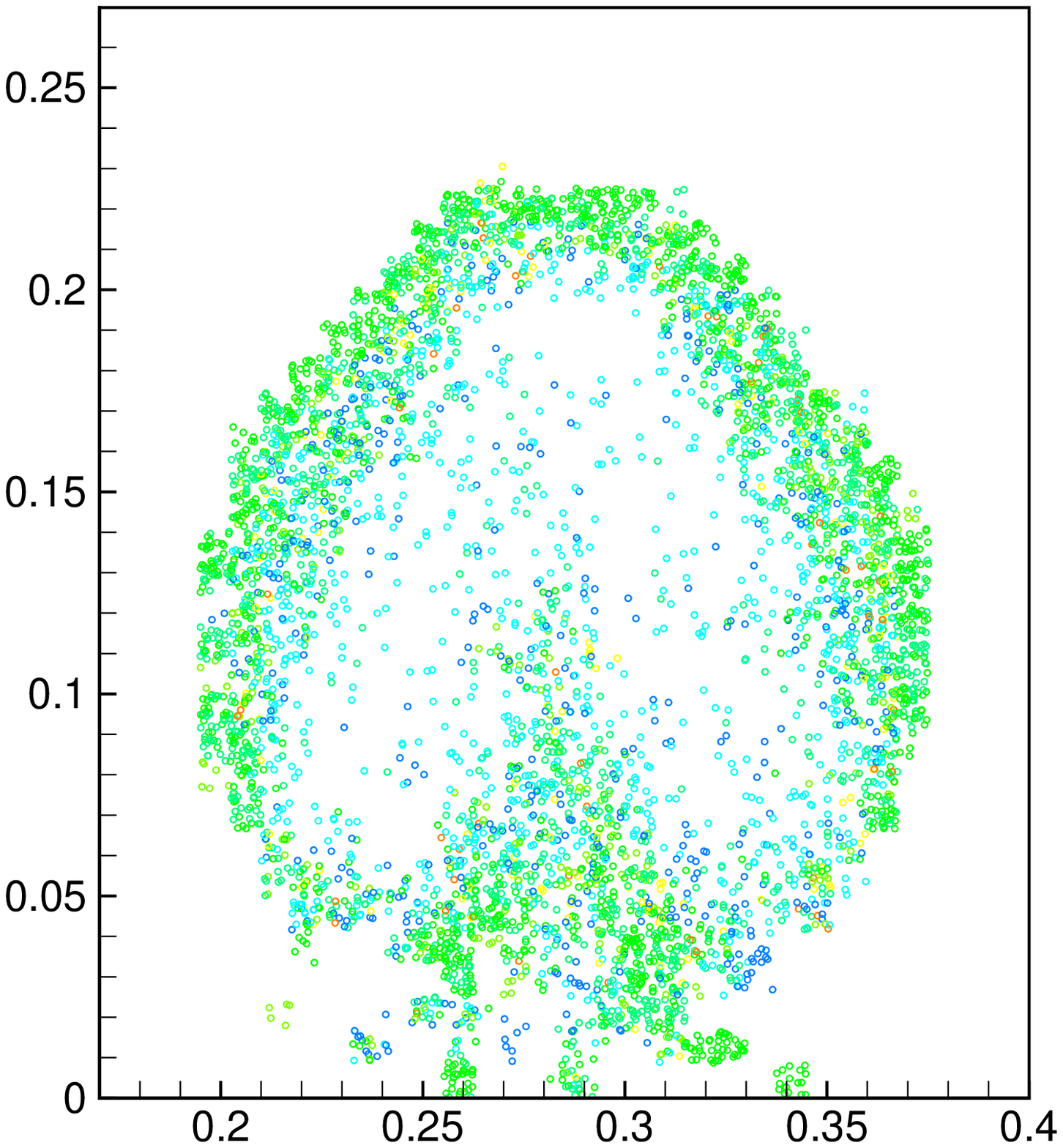}}
	\caption{Sampled particle for the solid particle phase in bubble formation process: from left to right are the results at time $0.05s$, $0.10s$, $0.15s$, and $0.18s$. The color shows the mass fraction of particle representation in UGKWP. The wave representation of solid particle phase in the dense particle zone is not shown here.}
	\label{bubble formation part in wp}
\end{figure}

The contours of apparent density of solid particle phase at different times are shown in Figure \ref{bubble formation apparent density}. The results show the typical process of bubble formation: initially, a small bubble occurs due to the jet with high velocity from the orifice; it becomes larger and larger in the evolution, and finally detaches the bottom boundary. During the process, the bubble shape is similar to an ellipse. The above process obtained by GKS-UGKWP agrees well with the observed phenomenon in the experiment \cite{Gasparticle-fluidized-single-bubble-nieuwland1996bubble}. To further quantitatively compare the bubble formation process with the experiment, the equivalent bubble diameter is calculated, which is defined as $D_e=\sqrt{4S/\pi}$. According to \cite{Gasparticle-fluidized-single-bubble-nieuwland1996bubble}, $S$ is the area of bubble obtained by the numerical simulation, defined as the area of $\epsilon_{s} < 0.15$. The equivalent bubble diameter obtained by GKS-UGKWP is presented in Figure \ref{bubble formation ed compare}, and it agrees well with the experiment measurement \cite{Gasparticle-fluidized-single-bubble-nieuwland1996bubble}, showing the accuracy and reliability of GKS-UGKWP. Besides, the sampled particles in UGKWP at different times are shown in Figure \ref{bubble formation part in wp}.
The original high-concentration solid particle bed is represented by wave, and isn't shown here. The sampled particles only appear in the non-equilibrium region, such as at the boundary between dense and dilute solid particle phase.
In addition, as the gas bubble becomes larger, more particles will emerge in UGKWP to capture the larger non-equilibrium zone with the penetration of solid particles in the gas bubble region.

\subsection{Particle clustering in fluidized bed}
\begin{figure}[htbp]
	\centering
	\subfigure{
		\includegraphics[height=5.0cm]{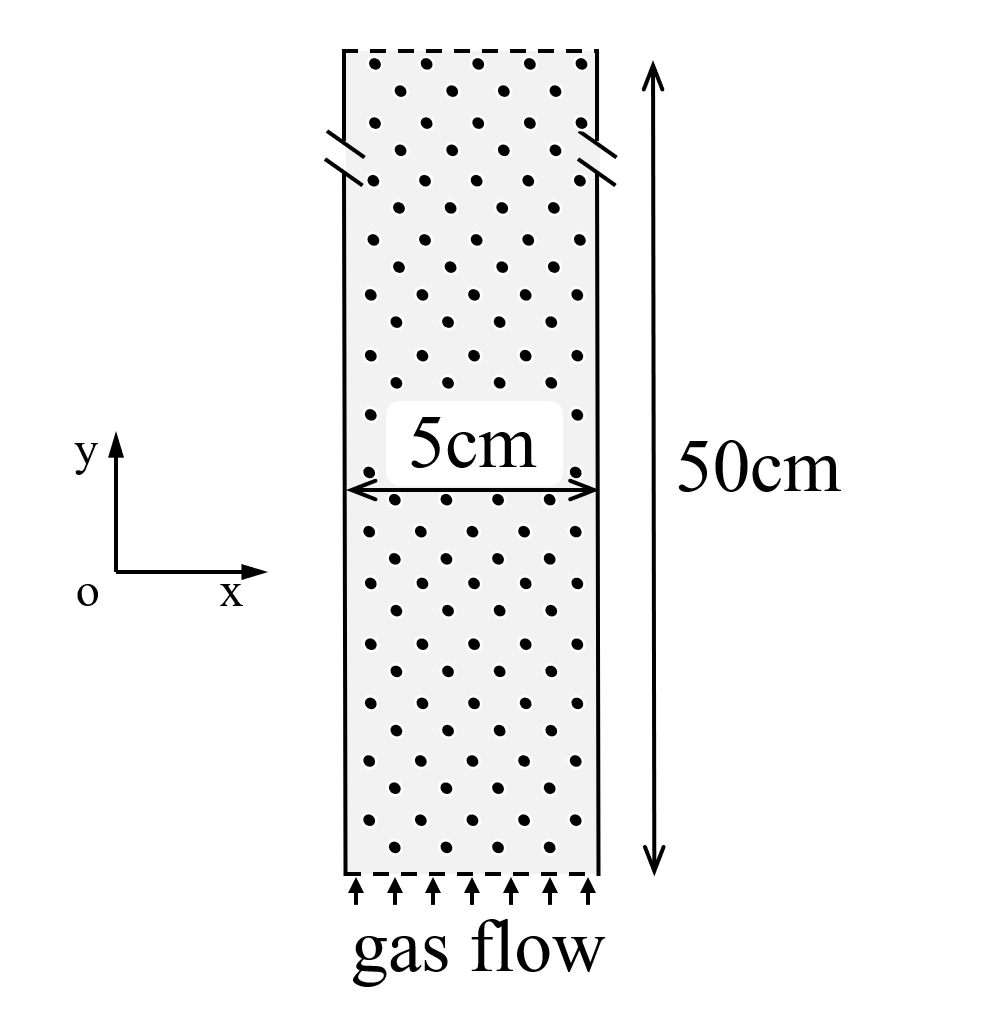}}
	\caption{Sketch of the vertical riser.}
	\label{Sketch of the vertical rise}	
\end{figure}

Particle clustering is a typical hydrodynamic phenomenon in circulating fluidized bed (CFB), and it has a significant influence on the evolution of gas-particle flow. In this section, GKS-UGKWP is used to calculate the CFB problem in \cite{Gasparticle-fluidized-circulating-helland2005numerical} and capture the particle clustering phenomenon. Figure \ref{Sketch of the vertical rise} presents the schematic diagram of the vertical riser in this problem. The computational domain $W\times H$ is $5cm\times50cm$ covered by $25\times250$ uniform rectangular mesh. Initially, the solid particles are distributed uniformly in the riser with the solid phase volume fraction 0.03 and zero velocity; the gas phase is in standard atmospheric condition, $\rho_g=1.2kg/m^3$, $p=1bar$, and zero velocity. The density and diameter of the solid particles in the riser are $2400kg/m^3$ and $133\mu m$ respectively. Initially, the air flows into the riser through bottom boundary with vertical velocity $V_g=1.0m/s$ and higher pressure approximated by $\Delta p = \epsilon_{s}\left(\rho_{s}-\rho_g\right)GH$. The solid particles are free to leave at the up boundary, and the escaped particles from the up boundary will be compensated back into the riser through the bottom boundary to maintain that the total mass of solid particles inside the riser is a constant in the whole simulation. For left and right walls, the slip and non-slip boundary conditions are employed for the solid phase and the gas phase respectively.

\begin{figure}[htbp]
	\centering
	\subfigure{
		\includegraphics[height=10.0cm]{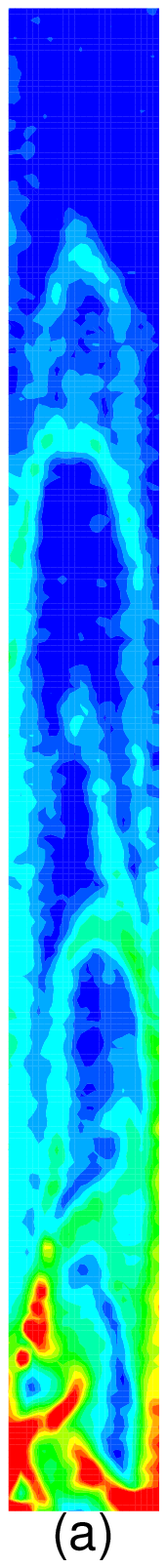}}
	\quad
	\subfigure{
		\includegraphics[height=10.0cm]{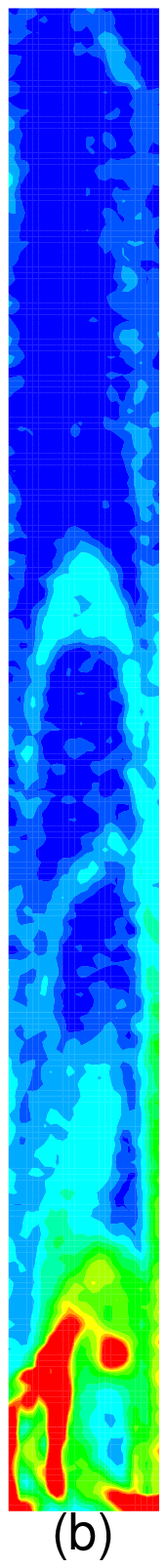}}
	\quad
	\subfigure{
		\includegraphics[height=10.0cm]{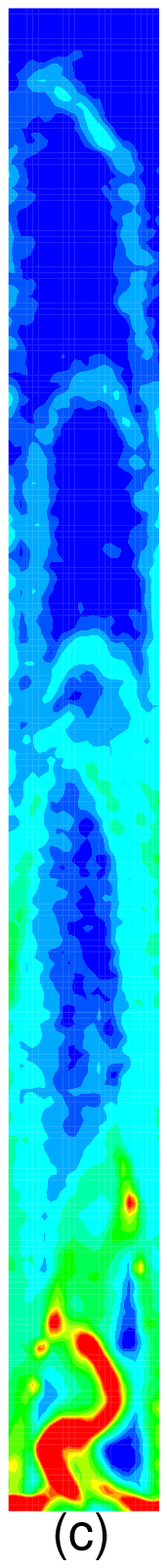}}
	\quad
	\subfigure{
		\includegraphics[height=10.0cm]{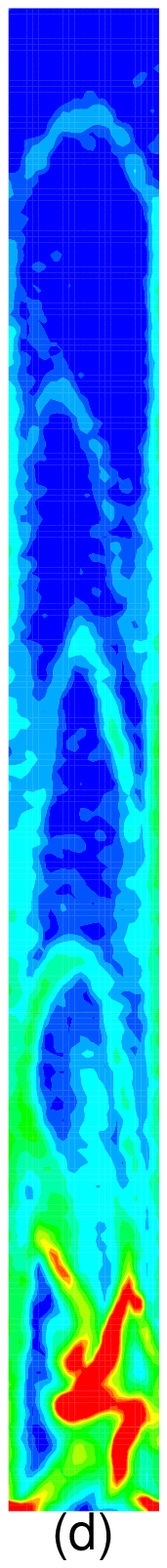}}
	\quad	
	\subfigure{
		\includegraphics[height=10.0cm]{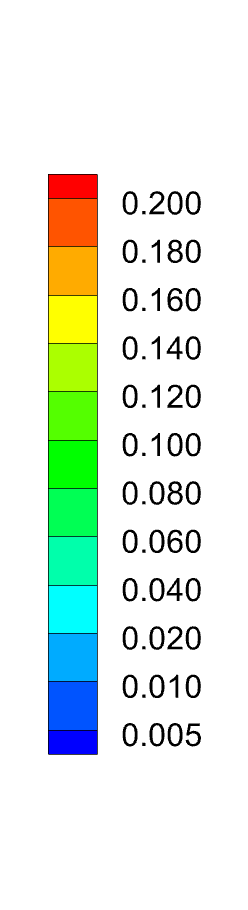}}
	\caption{The instantaneous snapshots of the distribution of solid phase volume fraction $\epsilon_s$ at different times: (a)$t=3.0s$, (b)$t=4.0s$, (c)$t=5.0s$, and (d)$t=6.0s$.}
	\label{vertical riser epsilons different times}	
\end{figure}

\begin{figure}[htbp]
	\centering
	\subfigure{
		\includegraphics[height=5.5cm]{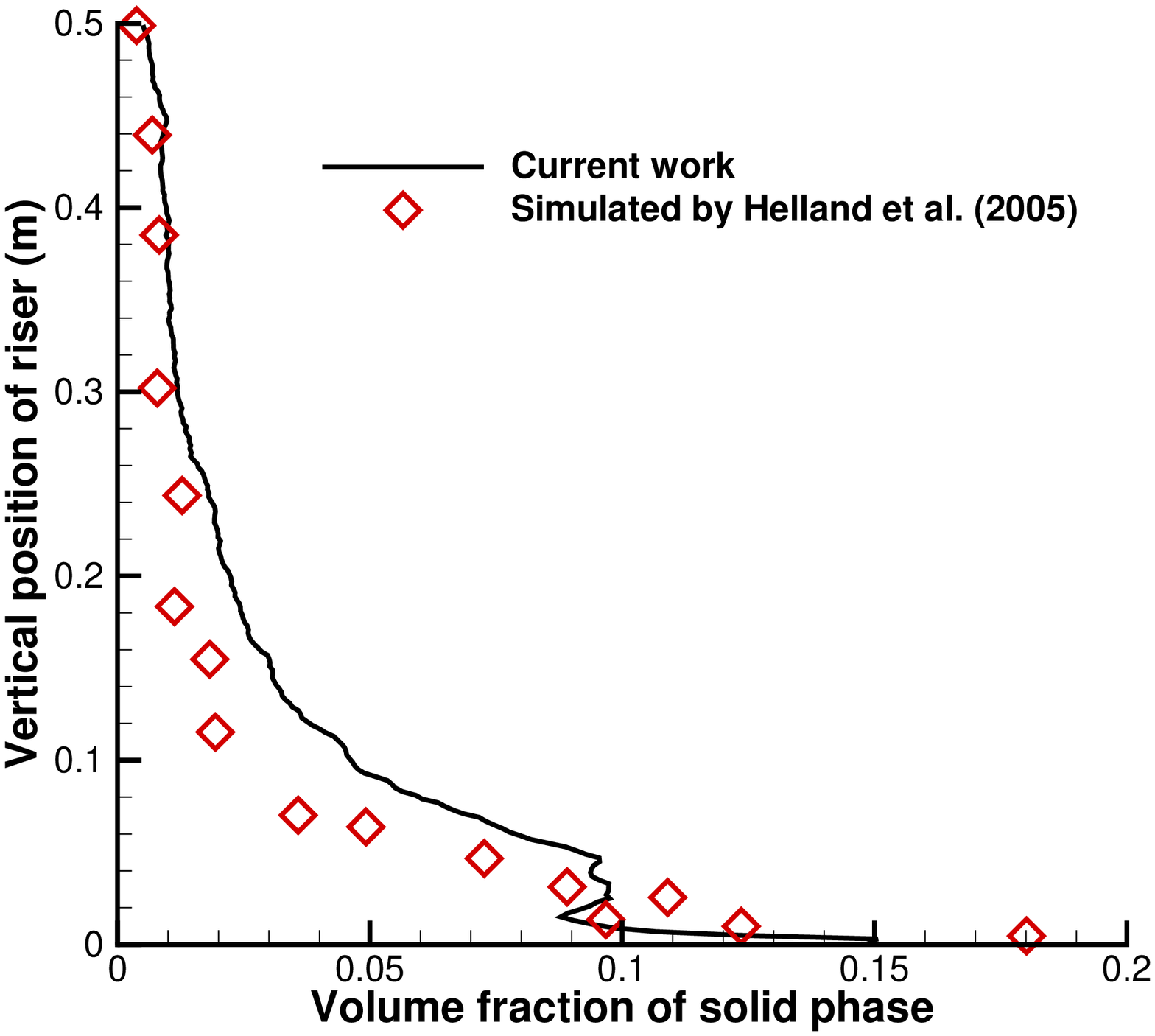}}
	\quad
		\subfigure{
		\includegraphics[height=5.5cm]{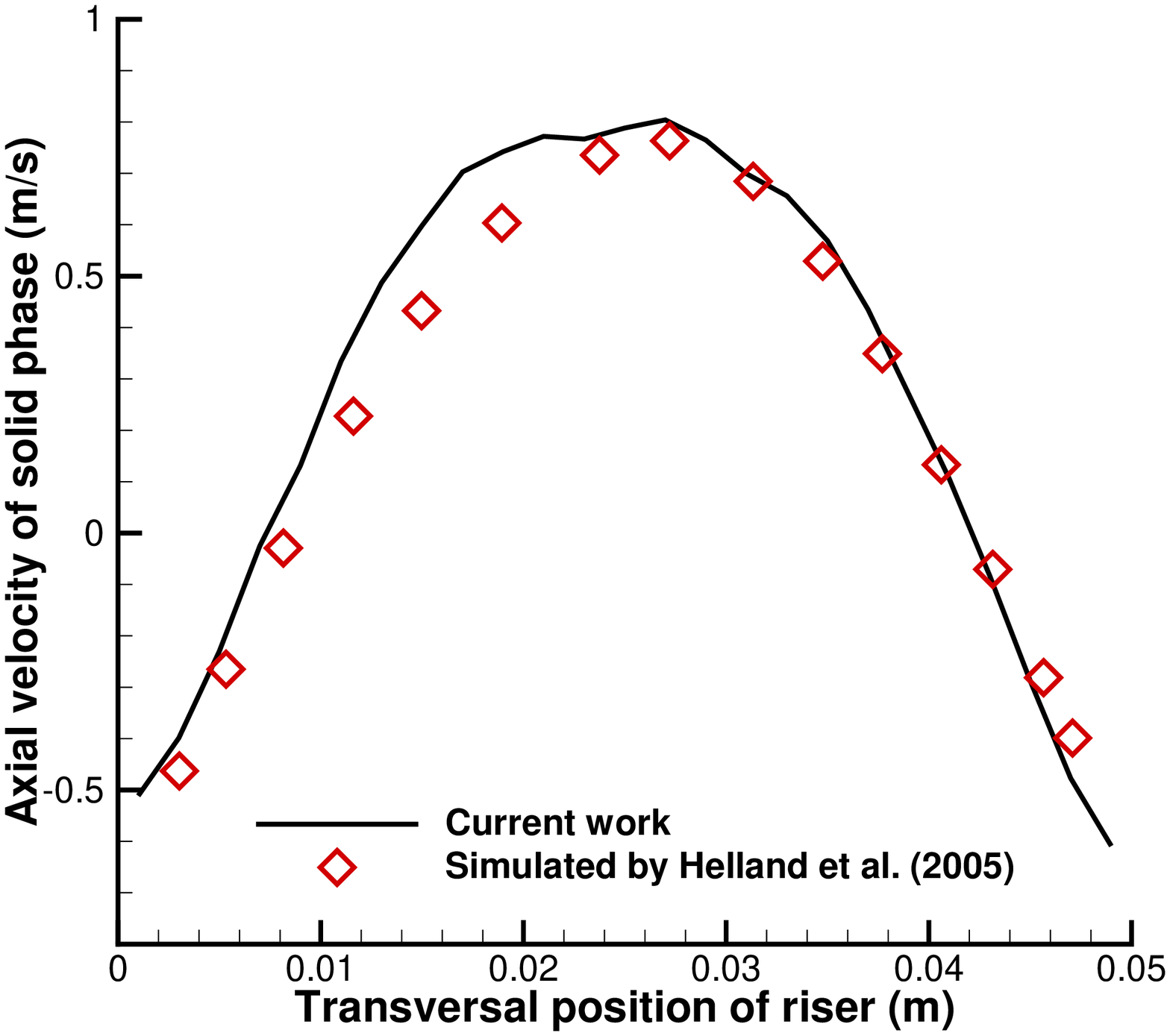}}
	\caption{Comparison with the numerical results by Eulerian-Lagrangian approach \cite{Gasparticle-fluidized-circulating-helland2005numerical}. Left: time-averaged solid phase volume fraction $\epsilon_s$ at different height. Right: transversal profile of the time-averaged solid phase velocity $v_s$ in the upper part of the riser.}
	\label{vertical riser results compare}
\end{figure}

The instantaneous snapshots of the distribution of solid volume fraction $\epsilon_s$ at different times are shown in Figure \ref{vertical riser epsilons different times}.
The results indicate that the typical heterogeneous structures in a circulating fluidized bed are captured clearly: axially it is dilute flow in the upper zone while dense flow in the bottom zone; solid particles aggregate into clusters in the riser; generally, solid particles and clusters are carried upward in the core zone by high-speed gas flow while dropping down mainly at the near-wall zone. All the above typical features are consistent with the previous observations in both numerical and experimental studies \cite{Gasparticle-fluidized-circulating-helland2005numerical}.
To further quantitatively analyze the results, the time-averaged profile is shown in Figure \ref{vertical riser results compare} and compared with the previous numerical results obtained by the Eulerian-Lagrangian approach \cite{Gasparticle-fluidized-circulating-helland2005numerical}. The profile of time-averaged $\epsilon_{s}$ at different riser height is shown in Figure \ref{vertical riser results compare}(a). The particle phase
has a lower concentration $0.01$ in the up zone, while a higher concentration $0.1$ in the zone near bottom boundary.
Figure \ref{vertical riser results compare}(b) presents the transversal profile of vertical velocity of particle flow $v_s$, which shows a parabolic shape, indicating solid particles move upward in the center region, while downward in the near-wall zone. Overall, the prediction given by GKS-UGKWP agrees well with the previous study by Eulerian-Lagrangian approach.

\section{Conclusion}
In this paper, GKS-UGKWP method is developed to study gas-particle two-phase flow with both dense and dilute solid particle concentration.
A drag force model for both dilute and dense particle flow is employed.
The  pressure model for inter-particle contacts and frictions is introduced, and it works for high solid particle concentration flow.
In addition, a flux limiting model is proposed to prevent the over-packing of the solid particle phase.
The non-conservative terms in the gas phase for accounting nozzle effect in momentum equation and $pDV$ work term in the energy equation, are added in the current scheme.
For the particulate flow at high concentration, the inter-particle collisions play significant roles in the evolution.
The inter-particle collision is included in the collision term of the kinetic equation for the particle phase to approach to the local equilibrium state. The current method can be used for particulate flow with a wide range of solid concentrations: from very dilute flow to dense one.

UGKWP is a multiscale method and is capable of capturing the multiscale transport of particulate flow efficiently by its coupled wave-particle formulation in the evolution process. At a small cell $Kn$ number in high particle concentration region, the intensive inter-particle collisions will drive the particle distribution to near equilibrium and is represented by wave component in UGKWP without particles. As a result, the EE two fluid approach can be recovered by UGKWP, the so-called coupled hydrodynamic equations for two phase flow.
While at large $Kn$ number for dilute particle  concentration, the inadequate inter-particle collision in UGKWP keeps the particle phase in non-equilibrium and its evolution is fully determined by the particle transport. The EL approach for the two phase flow is obtained by UGKWP automatically in the dilute particle concentration region. At an intermediate $Kn$ number, both wave and particle in UGKWP play roles in the evolution, and the number of sampled particles is determined by the local degree of flow non-equilibrium, which ensures a smooth and consistent transition in different flow regimes.

The proposed GKS-UGKWP for the gas-particle system is tested by a series of two-phase problems. The interaction of shock wave with solid particle layer in a channel is simulated, and the numerical results agree well with the previous study by EL approach. In the horizontal pneumatic conveying problem, typical flow patterns observed in the experiment for both low and high solid concentrations are well captured by GKS-UGKWP.
The bubble formation through a particle bed is well captured by the proposed method and the bubble shape and size agree well with the experiment measurements. Also in the circulating fluidized bed case, the particle clustering phenomenon and the corresponding heterogeneous structures are well captured by GKS-UGKWP. These results validate the accuracy and reliability of GKS-UGKWP for the simulation of gas-particle two-phase flow.

\section*{Acknowledgements}
The current research is supported by National Numerical Windtunnel project, National
Science Foundation of China (11772281, 91852114,12172316), Hong Kong research grant council 16208021, and Department of Science and Technology of Guangdong Province (Grant No.2020B1212030001).

\bibliographystyle{plain}%
\bibliography{jixingbib1}
\end{document}